\documentclass{article}
\usepackage{jheppub}
\usepackage{physics}

\newcommand\myrenyi[0]{R\'enyi }
\begin{document}

\title{The entanglement entropy of typical pure states and replica wormholes}

\author{Erez Y.~Urbach}
\affiliation{Department of Particle Physics and Astrophysics, Weizmann Institute of Science, Rehovot, Israel}
\emailAdd{erez.urbach@weizmann.ac.il}
\abstract{In a 1+1 dimensional QFT on a circle, we consider the von Neumann entanglement entropy of an interval for typical pure states. As a function of the interval size, we expect a Page curve in the entropy. We employ a specific ensemble average of pure states, and show how to write the ensemble-averaged \myrenyi entropy as a path integral on a singular replicated geometry. Assuming that the QFT is a conformal field theory with a gravitational dual, we then use the holographic dictionary to obtain the Page curve. For short intervals the thermal saddle is dominant. For large intervals (larger than half of the circle size), the dominant saddle connects the replicas in a non-trivial way using the singular boundary geometry. The result extends the `island conjecture' to a non-evaporating setting.}
\maketitle

\section{Introduction}
	One of the main recent developments in the study of quantum gravity are replica wormholes: non-trivial gravitational solutions connecting the replicas in the quantum gravitational replica-trick \cite{Penington:2019kki,Almheiri:2019qdq}. These solutions give rise to a refinement of the gravitational von Neumann entropy formula called the island conjecture \cite{Almheiri:2019psf,Penington:2019npb,Almheiri:2020cfm}. In the context of evaporating black holes, these solutions are dominant at large times, leading to the Page-curve of the black hole radiation. 

\begin{figure}
	\centering
	\includegraphics[width=.5\linewidth]{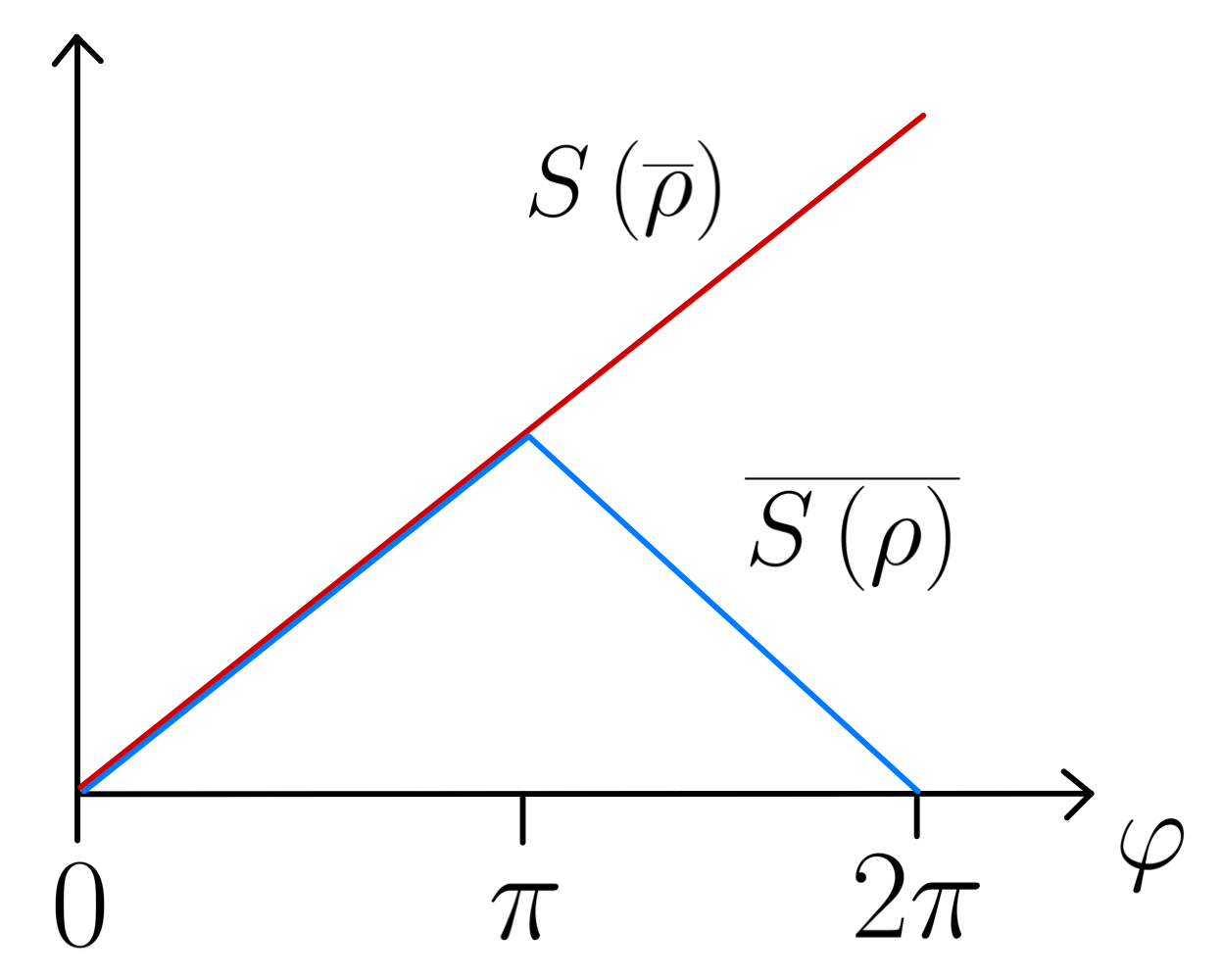}
	\caption{In $1+1$ dimensions, we consider a theory on a circle with periodicity $2\pi$. The graph is a schematic picture for the von Neumann entropy on the interval $A=[0,\varphi]$. The thermal result (in red) is calculated by first averaging the pure density matrix and then taking the von Neumann entropy. The pure `Page-like' result can be obtained by averaging over the pure von Neumann entropies (in blue).
	\label{fig:page_curv}
	}
\end{figure}

	In this letter, we study a purely gravitational system on $AdS_{d+1}$ with no evaporation or dynamics. In the dual field theory on spatial $S^{d-1}\times \mathbb{R}$ we ask for the von Neumann entropy $S_{vN}(A)$, $A$ a region of $S^{d-1}$, of a typical high-energy pure state. Holographically this quantity describes the spatial encoding of a typical $d+1$-dimensional AdS black hole microstate in the dual boundary theory. 
	For a typical state, we expect $S_{vN}(A)$ to follow a Page curve in $A$ (see figure \ref{fig:page_curv}). In fact, this is exactly the toy model used by Page for black hole evaporation \cite{Page:1993df,Page:1993wv}. By employing an ensemble average over microstates we show how to get this curve using the standard holographic dictionary of Euclidean gravity. We find a phenomenon similar to replica wormholes, but without conjecturing their existence in the quantum gravity path integral. Instead, the non-trivial replica topologies are imposed by the field theory calculation, from the purity of ensemble states. The result is a generalization of the standard thermal RT formula \cite{Azeyanagi:2007bj} for the typical pure state, where now the RT surface has a weaker homology constraint \cite{Haehl:2014zoa}.
	
	A closely related quantity is the thermal density matrix. Compared to an energetic pure state, the von Neumann entropy of the thermal density matrix is a  more controlled quantity. One reason is that thermal entropies can be written in field theory as a Euclidean path integral over some geometry. The holographic dictionary can then be used to map these quantities to classical observables in the dual gravitational theory \cite{Lewkowycz:2013nqa}. In this way, the thermodynamic entropy $S_{th}(\beta)$ was found by Gibbons and Hawking \cite{Gibbons:1976ue} to be proportional to the dual black hole horizon area. Similarly, the thermal von Neumann entropy $S_{vN}(A)$ is given by the RT formula in the background of a stationary AdS black hole \cite{Ryu:2006bv,Nishioka:2009un,Ryu:2006ef}.
	However the von Neumann entropy of a generic high-energy microstate can't be written in such a simple path integral but typically requires $O(1/G_N)$ operator insertions. While being practically impossible to calculate exactly on the field theory side, 
	it is unclear to what extent semiclassical analysis applies to a specific black hole microstate.\footnote{For the RT result of non-generic black holes created by a collapse see \cite{Takayanagi:2010wp}. For the RT result for (pure) CFT B-states see \cite{Takayanagi:2011zk,Hartman:2013qma}.}

	Instead of looking at a specific microstate, we propose an ensemble of pure states, weighted by a Boltzmann-like factor parametrized by $\beta$. Taking the ensemble average over the different microstate entropies ``$\overline{S_{vN}(A)}$" allows us to write it as a path integral similar to the thermal case, albeit over a singular geometry. The resulting averaged entropy claims to capture the von Neumann entropy of a typical microstate with a given energy $E\sim 1/\beta$.
	The singularity of the integral is a result of the pure state being the same one on all the different replicas in the \myrenyi entropy calculation. The singular behavior allows for several equivalent geometrical interpretations of the field theory replica path integral. One geometry resembles the thermal calculation, while a second geometry connects all replicas together (see figure \ref{fig:topologies}). The field theory path integrals over each of the geometries are equal mathematically.

	Writing the averaged entropy as a path integral, we can employ the holographic dictionary to find its value. We propose that a holographic calculation can be made by taking all the field-theory geometries as possible asymptotic boundaries for the gravitational path integral. Each boundary geometry gives (in the large $1/G_N$ saddle-point approximation) a different saddle (see figure \ref{fig:pure_RT}). Whenever dominant, the saddle with the standard thermal boundary geometry gives the thermal RT result $\overline{S_{vN}(A)} = \frac{\text{Area}(X_1)}{4G_N}$. Here $X_1$ is the RT surface in the background of an AdS black hole \cite{Azeyanagi:2007bj}. Notably, the RT surface $X_1$ is homologous to the (asymptotic) entanglement region $A$. 
	The second boundary geometry has a corresponding gravitational saddle which smoothly connects the replicas non-trivially (also) in the bulk. The resulted contribution of this saddle to the von Neumann entropy (when dominant) can be written as $\overline{S_{vN}(A)} = \frac{\text{Area}(X_2)}{4G_N}$. $X_2$ is another extremal surface of the same background geometry as $X_1$, but one homologous to the complement $A^c$. The bulk entangling surface can be understood in this case as an `island' covering the horizon.

	The final result is a refinement of the RT formula for the case of typical pure states, where we allow the RT surface $X$ to be homologous to either $A$ or its complement $A^c$.\footnote{For non-typical pure states with a known geometric dual, it should possible to retain the homology constraint. One example of that is the non-typical collapsed black hole state, where the homology constraint can be kept due to the past geometry \cite{Takayanagi:2010wp}. Another is a CFT `B-state' in which non-trivial surfaces are homologically allowed due to an end of the world brane \cite{Hartman:2013qma}.}
	At small $A$ the thermal saddle dominates, giving an agreement of the typical pure state entropy with the thermal entropy. This result is supported by expectations from eigenstate-thermalization-hypothesis (ETH) calculations \cite{Goldstein:2005aib,Dymarsky:2016ntg,Fujita:2017pju} and large central charge expansions \cite{He:2017vyf}.
	As a function of the entanglement region $A$, a Page-curve occurs as the non-trivial saddle becomes dominant over the thermal saddle for large enough regions (figure \ref{fig:phase_transition}). The non-trivial saddle thus ensures the `purity' of the result $\overline{S_{vN}(A)}=\overline{S_{vN}(A^c)}$. Therefore, both in the field theory and in gravity, the non-trivial geometry of the replica calculation is directly related to the purity of the state.

	We start in section \ref{sec:ensemble} by describing the ensemble of pure-states we will use to calculate the averaged entropy. The calculation of the averaged pure state entropy is best understood as a generalization of the thermal entropy calculation. We, therefore, begin in section \ref{sec:thermal} by briefly reviewing the von Neumann entropy calculation for the thermal density matrix. 
	We explain how in field theories the von Neumann entropy can be written as a replica path integral, and (following \cite{Lewkowycz:2013nqa}) how computing the path integral using the holographic dictionary gives the RT formula. Finally, in section \ref{sec:typical} we show how a similar calculation can be made studying the von Neumann entropy averaged over the ensemble of pure states defined in section \ref{sec:ensemble}. In particular, we show how for any field theory the averaged \myrenyi entropy can be written as a path integral \eqref{eq:av_renyi_path}. Using the holographic dictionary we find our proposed refinement of the RT formula for typical pure states \eqref{eq:typical_RT_proposal}.

\section{The ensemble of states} \label{sec:ensemble}
We will focus here on Euclidean two-dimensional field theories on $S^1_{2\pi}\times \mathbb{R}$ but everything can be generalized to theories in a general dimension $d$ on $S^{d-1}\times \mathbb{R}$. We denote the fields in the theory collectively by $\phi(\theta,\tau)$. In the main text, we assume $\phi$ is a scalar. We believe the calculations described can be extended, with suitable adjustments, to fermions and gauge fields. Specifically, they are expected to be generalizable to known holographic CFTs such as $\mathcal{N}=4$ SYM on $S^{3}\times \mathbb{R}$.

We start by defining our ensemble of pure states $\left\{\ket{\psi_\alpha}\right\}$, $\alpha$ is the ensemble index. First, we choose a set of commuting operators that we denote collectively $\mathcal{O}$ and its eigenbasis $\ket{\alpha}$ with $\mathcal O \ket{\alpha} = \lambda_{\alpha} \ket{\alpha}$, which we assume to uniquely identify the state. Our ensemble of pure states is the Euclidean evolution of this basis
\begin{equation}\label{eq:ensemble_state}
	\ket{\psi_\alpha} = e^{-\frac{\beta}{2} H} \ket{\alpha},
\end{equation}
for all $\alpha$.\footnote{The ensemble is very similar to that of \cite{Sugiura:2012,Fujita:2017pju}.} Over the following sections, our favorite choice for the operators $\mathcal O$ would be all the field operators 
$\left\{\hat\phi(\theta) \right\}$, which correspond to the field state basis $\ket{\alpha}\equiv \ket{\phi_\alpha}$ for every field configuration $\phi_\alpha(\theta)$.\footnote{Given a gauge field $A_\mu(\theta,\tau)$, similar basis can be formed.  The basis include $\ket{\alpha}=\ket{(A_\alpha)_\theta}$ for every spatial gauge field $(A_{\alpha})_\theta(\theta)$, up to gauge transformations $(A_{\alpha})_\theta(\theta) \sim (A_{\alpha})_\theta(\theta) +\partial_\theta f $.} 
For this choice of basis the ensemble states (sandwiched with a field-state $\ket{\phi_0}$) can be written as a Euclidean path integral
\begin{equation}
	\bra{\phi_0}\ket{\psi_\alpha} = \int D\phi {\bigg\vert}_{\phi(\theta,-\frac{\beta}{2})=\phi_\alpha(\theta)}^{\phi(\theta,0)=\phi_0(\theta)} e^{- S[\phi]} = 
	\vcenter{\hbox{\includegraphics[width=.3\linewidth]{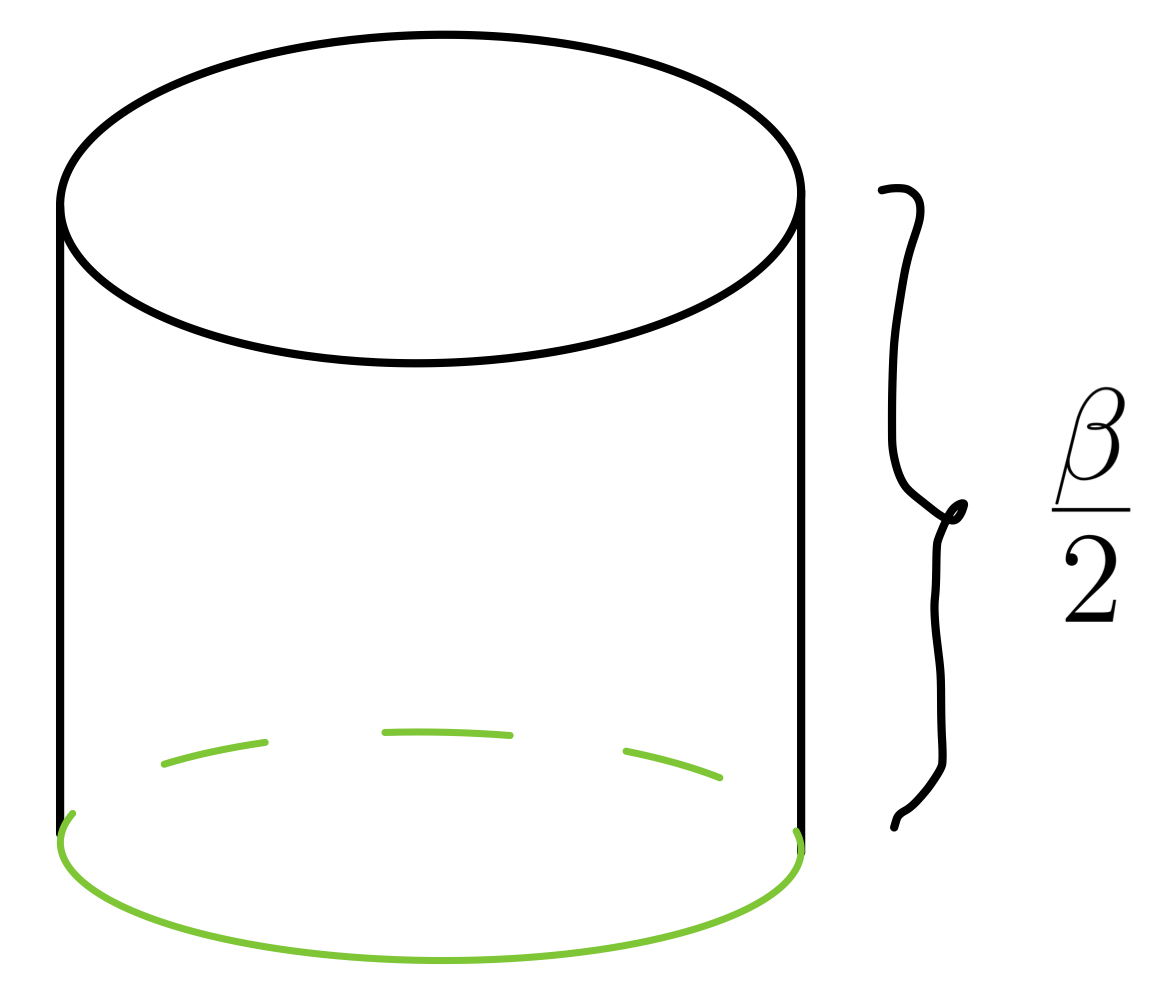}}}.
\end{equation}
For any other basis a similar path integral can be written, only the identification (the green circle) at $\tau=-\frac{\beta}{2}$ would be $\delta(\mathcal{O}-\lambda_\alpha)$. The role of the Euclidean evolution is to select on average states with energy $E\sim 1/\beta$. One way to see it is to decompose the ensemble states into energy eigenstates $\ket{\alpha}=\sum_n c_n \ket{E_n}$. The energy of the state $\ket{\psi_\alpha}$ is thus
\begin{equation}
	\frac{\bra{\psi_\alpha}H\ket{\psi_\alpha}}{\bra{\psi_\alpha}\ket{\psi_\alpha}} = \frac{\sum_n |c_n|^2 E_n e^{-\beta E_n}}{\sum_n |c_n|^2 e^{-\beta E_n}}
\end{equation}
For generic operators $\mathcal{O}$ we expect the $c_n$'s to be random, and each state $\ket{\psi_{\alpha}}$ should resemble a typical state in the canonical ensemble. Note that the energy basis will completely localize the $c_n$s and won't give the necessary Boltzmann suppression. One can try to fix it by using a microcanonical ensemble around a small energy window, but this option has other issues we will discuss later (see footnote \ref{foot:typicality}).

For a given element of the ensemble, we can define the unnormalized density matrix $\rho_\alpha=\ket{\psi_\alpha}\bra{\psi_\alpha}$. We separate the spatial circle into two regions: $A=[0,\varphi]$ and its complement $A^c = [\varphi,2\pi]$. The reduced density matrix is $\rho_\alpha(\varphi)=\frac{1}{\text{Tr}\rho_\alpha}\text{Tr}_{A^c} \rho_\alpha$, which can be drawn as
\begin{equation}\label{eq:density_matrix}
	\text{Tr}_{A^c} \rho_\alpha= 
	\vcenter{\hbox{\includegraphics[width=.25\linewidth]{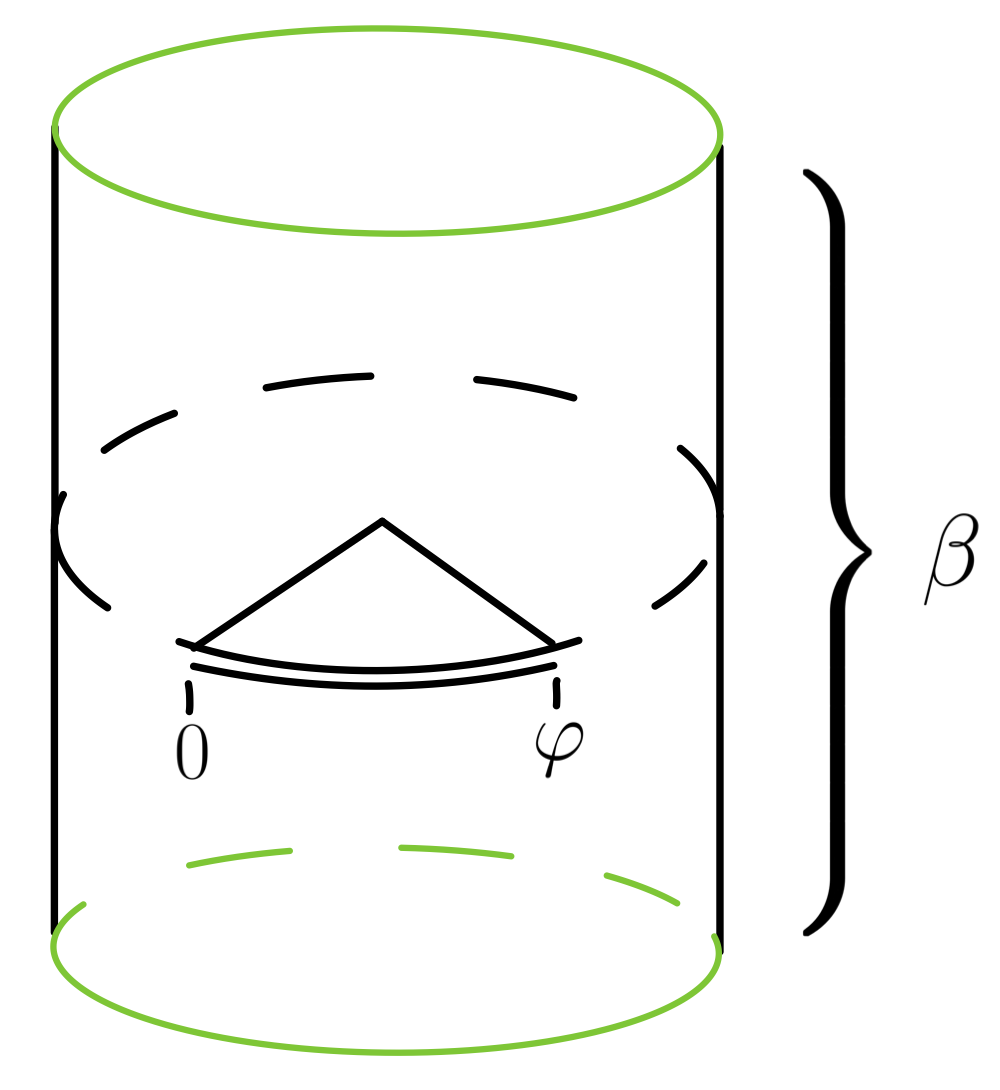}}}
	=
	\vcenter{\hbox{\includegraphics[width=.25\linewidth]{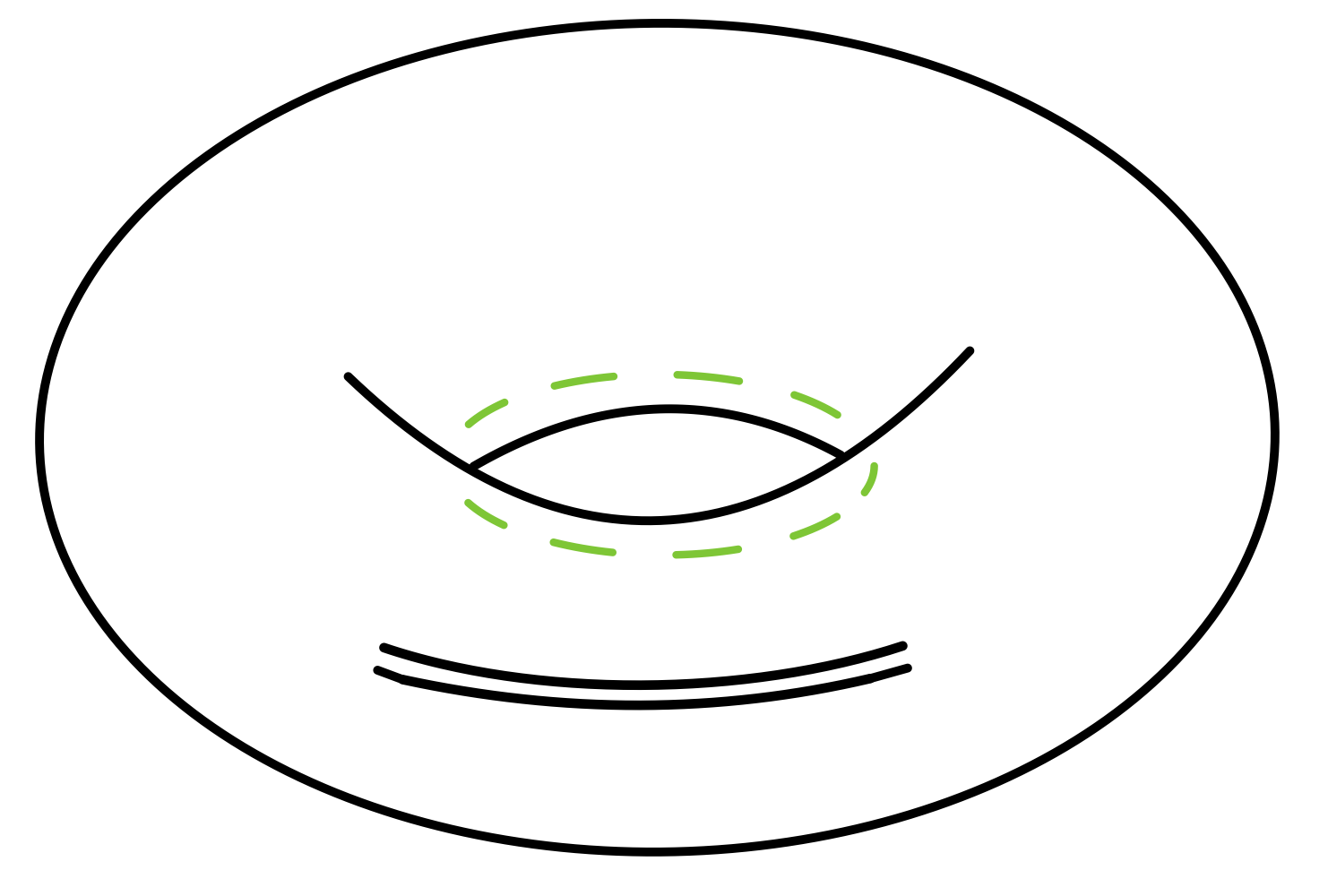}}}.
\end{equation}
The green circles represent the $\alpha$ identification at $\tau=\pm \frac{\beta}{2}$. The RHS represents the same calculation only in a shape of a torus $S^1_{2\pi}\times S^1_{\beta}$. The horizontal circle is the spatial $\theta$ direction with periodicity $2\pi$. The vertical circle is the euclidean time direction $\tau$ with length $\beta$. The cut is on the spatial interval $[0,\varphi]$ and euclidean time $\tau=0$. The green spatial circle covers all of $\tau=\pm\frac{\beta}{2}$.

We want to calculate ensemble-averaged quantities, which we will denote with a bar $\overline{(...)}$. Given a quantity $W_\alpha$ calculated on each ensemble state $\ket{\psi_\alpha}$, the averaged quantity can be written formally as $\overline{W}=\frac{1}{\sum_\alpha 1} \sum_\alpha W_\alpha$. For discrete bases, this sum has a well-defined meaning. For continuous bases, like the field basis $\ket{\phi_\alpha}$, we need to specify a measure on the formal sum, and the value of $\overline{W}$ will depend on that choice. A similar question arises when one needs to define the trace at a given basis, where formally $\text{Tr}(...)=\sum_{\alpha} \bra{\alpha} ... \ket{\alpha}$. Up to an overall constant, we will use the same measure used for the trace. In the case of the field basis $\ket{\phi_\alpha}$, this amounts to a path integral $``\sum_\alpha"=\int D\phi$.

\section{The thermal mixed state}\label{sec:thermal}
\subsection{QFT side}
As a first step, we would like to find the entanglement-entropy of the averaged density matrix $S_{vN}(\overline{\rho}(\varphi))$. We define the ensemble-averaged density matrix $\bar \rho$ by the (normalized) statistical mixture of all the individual $\rho_\alpha$ in the ensemble. The result is simply the thermal density matrix
\begin{equation} \label{eq:av_rho_thermal}
\begin{split}
	\overline \rho & \equiv \frac{1}{Z}\sum_\alpha \rho_\alpha = \frac{1}{Z}\sum_{\alpha} e^{-\frac{\beta}{2} H}\ket{\alpha}\bra{\alpha} e^{-\frac{\beta}{2} H} \\
	& = \frac{1}{Z} e^{-\beta H} = \rho_\text{thermal}.
\end{split}
\end{equation}
The normalization is $Z=\sum_\alpha \bra{\alpha}e^{-\beta H}\ket{\alpha}$. Note that the result is independent of the basis we chose (or the operators $\mathcal{O}$).\footnote{One can imagine a different type of averaged density matrix, where the mixture is between the normalized $\rho_\alpha$: $\overline{\rho}' = \frac{1}{\mathcal{N}}\sum_{\alpha} \frac{e^{-\frac{\beta}{2} H} \ket{\alpha}\bra{\alpha}e^{-\frac{\beta}{2} H} }{\bra{\alpha}e^{-\beta H}\ket{\alpha}}$ ($\mathcal{N}$ is the formal normalization). We didn't study this expression, but we believe the holographic calculation to be equal to the thermal case.}
In terms of path integrals, 
we simply replaced the boundary conditions by a geometrical identification as in the right-hand side of \eqref{eq:density_matrix}, but without the green circle.
Finally, we define the reduced averaged density matrix as $\overline{\rho}(\varphi)=\text{Tr}_{A^c} \rho_\text{thermal}$.
In the rest of the section, we briefly explain how to find the entanglement-entropy for $\bar\rho=\rho_\text{thermal}$ in field theory and using holography.
Although these results are well known in the literature they will help us motivate the calculation of the next section, for the pure typical state.

For every density matrix $\rho$, its von Neumann or entanglement entropy $S_{vN}(\rho)$ can be written using the \myrenyi entropy $S_{n}(\rho)$:
\begin{align}
	S_{vN}(\rho)&\equiv-\text{Tr}(\rho \log \rho)=\lim_{n\rightarrow 1} S_n(\rho), \label{eq:EE_def}\\
	S_n(\rho) &\equiv \frac{1}{1-n} \log \text{Tr} \rho^n. \label{eq:renyi_def}
\end{align}
Using \eqref{eq:av_rho_thermal}, we can write $\text{Tr}_A \overline{\rho}(\varphi)^n=\frac{Z_n}{Z_1^n}$. Here $Z_1$ is the partition-function on $\mathcal{M}_1=S^1_{2\pi}\times S^1_\beta$, and $Z_n$ is the partition-function over the $n$-sheeted torus we call $\mathcal{M}_n$.
To define it, start with $n$ copies of the torus  $\left(\mathcal{M}_1\right)^n=\left(S^1_{2\pi}\times S^1_\beta\right)^n$ and denote the intervals of the i'th replica $A_\pm^{(i)} = [0,\varphi]\times \{0^\pm\}$, $i=1,...,n$. $\mathcal{M}_n$ is defined by cutting these intervals and gluing them back by $A_+^i = A_-^{i+1}$ for $i=1,...,n-1$, and $A_+^n = A_-^{1}$. We can draw the result by
\begin{equation}
	Z_n= 
	\vcenter{\hbox{\includegraphics[width=.5\linewidth]{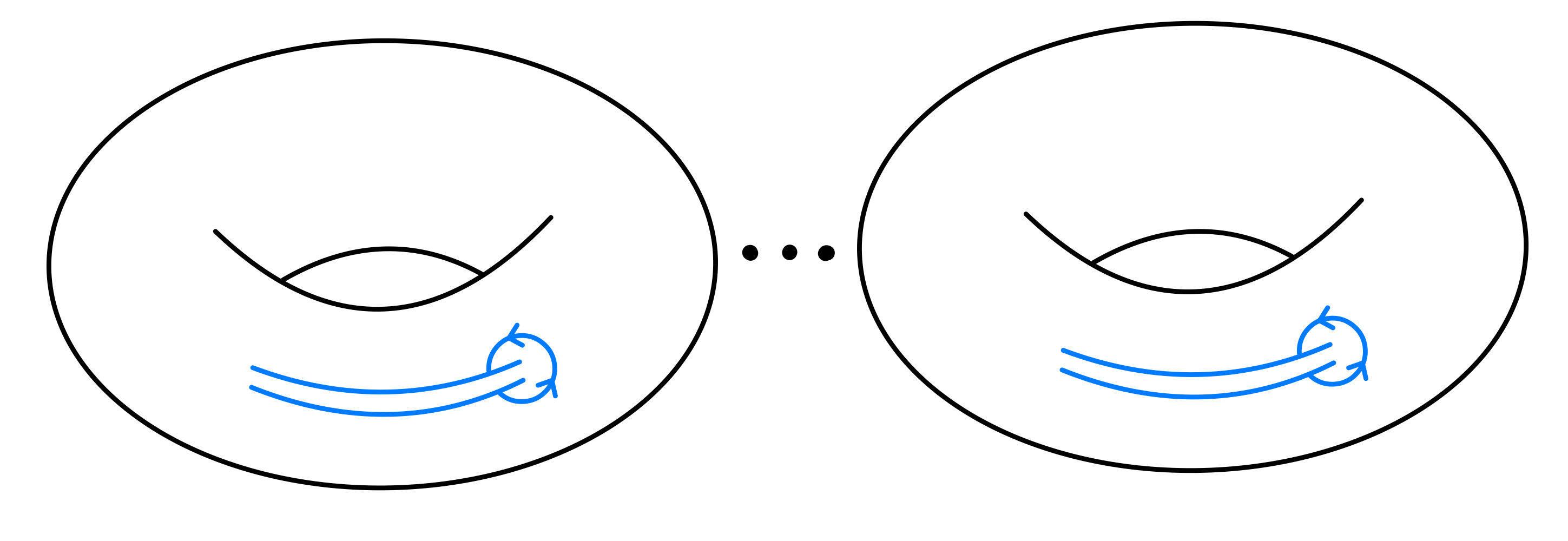}}}.
\end{equation}
Explicitly in terms of path integrals we can write
\begin{equation}\label{eq:thermal_renyi}
	S_n\left(\overline{\rho}(\varphi)\right) =
	\frac{1}{1-n}
	\log \frac{
		\int_{\mathcal{M}_n} D\phi \ e^{-S[\phi]}
		}{
		\int_{\left(\mathcal{M}_1\right)^n} D\phi \ e^{-S[\phi]}
	}.
\end{equation}

The behavior of the thermal von Neumann entropy $S_{vN}\left(\overline{\rho}(\varphi)\right)$ is known at several regimes. For low temperatures $\beta\gg 1$ the ensemble localize to the ground-state, and the result on $S^1$ is universal (up to a cutoff-dependent additive constant) \cite{Holzhey:1994we,Calabrese:2004eu,Ryu:2006ef}
\begin{equation}\label{eq:vac_ee}
	S_{vN}\left(\overline{\rho}(\varphi)\right) = \frac{c}{3} \log\left(\frac{2}{a} \sin\left(\frac{\varphi}{2}\right)\right),
\end{equation}
$a$ being the UV cutoff. We are interested instead in the high-temperature limit $\beta\ll 1$. At least for $\beta\ll\varphi\ll 1$ one can use the universal finite temperature limit on the line  \cite{Calabrese:2004eu,Ryu:2006ef}
\begin{equation} \label{eq:thermal_ee}
	S_{vN}\left(\overline{\rho}(\varphi)\right) \approx \frac{c}{6} \frac{\varphi}{\beta}.
\end{equation}
On general grounds, we expect the thermal entropy $S_{vN}\left(\overline{\rho}(\varphi)\right)$ to increase with $\varphi$ (at least for small enough $\varphi$), and to reach the thermodynamic entropy of the theory $S_{th}(\beta)$ at $\varphi=2\pi$. Note that unlike the individual $\rho_\alpha(\varphi)$, $\overline \rho$ is not pure, and so we don't expect $S_{vN}\left(\overline{\rho}(\varphi)\right)\ne S_{vN}\left(\overline{\rho}(2\pi-\varphi)\right)$ (see figure \ref{fig:page_curv}).

We would like to stress a trivial point. At the denominator of \eqref{eq:thermal_renyi} for example, we took the path integral over $\mathcal{M}_1=S^1_{2\pi}\times S^1_\beta$. The reason is that the normalization of the thermal density matrix is the trace $Z_1 = \text{Tr} e^{-\beta H}$. Using the path integral formalism, the trace can be written as a path integral on $S^1_{2\pi}\times [-\frac{\beta}{2},\frac{\beta}{2}]$ with periodic identification
\begin{equation}\label{eq:trace_canonical}
	Z_1 = \int_{S^1_{2\pi}\times \left[-\frac{\beta}{2},\frac{\beta}{2}\right]} D\phi{\bigg\vert}_{\phi\left(\theta,-\frac{\beta}{2}\right)=\phi\left(\theta,\frac{\beta}{2}\right)} e^{-S[\phi]}.
\end{equation}
Note that this is not directly the same as the path integral on the torus. In \eqref{eq:trace_canonical} only the field is identified at the two ends, whereas for the torus we assume only smooth configurations of the field. More concretely, in \eqref{eq:trace_canonical} a derivative discontinuity at $\tau=\pm\frac{\beta}{2}$ cost no action, unlike on the torus. The reason the two computations do agree is that smooth configurations on the torus can approximate functions with discontinuous derivative arbitrarily close to $\tau=\pm\frac{\beta}{2}$. We checked these claims explicitly for the free scalar and $n=2$ in appendix \ref{app:ident}. These comments also apply to the numerator of \eqref{eq:thermal_renyi}, where the path integral is over the $n$-sheeted torus $\mathcal{M}_n$. Although we discuss the field basis, similar statements should hold for other bases as well.

\subsection{Gravity side}\label{sec:termal_GR}
Holographically the calculation of the thermal entropy $S_{vN}\left(\overline{\rho}(\varphi)\right)$ at high temperatures corresponds to the study of RT surfaces in the background of an Euclidean AdS black hole \cite{Hawking:1982dh,Witten:1998zw,Azeyanagi:2007bj}. Below we follow  \cite{Lewkowycz:2013nqa,Ryu:2006bv} and emphasize the main ingredients leading to the RT formula.

The main observation of  \cite{Lewkowycz:2013nqa} is that both terms in the \myrenyi entropy \eqref{eq:thermal_renyi} can be translated by the AdS/CFT dictionary to concrete gravitational calculations
\begin{equation}\label{eq:renyi_grav}
	S_n\left(\overline{\rho}(\varphi)\right) =
	\frac{1}{n-1}
	\left( I_n(\varphi) - n\, I_1\right).
\end{equation}
Here $I_1$ is the gravitational on-shell action of the solution with a boundary of a torus $\mathcal{M}_1$, and $I_n(\varphi)$ with a boundary of $\mathcal{M}_{n}$. We stress again that the asymptotic topology of a torus (and not a cylinder) was due to the smoothness assumption we were allowed to take.
We are interested in the $\beta\ll 1$ limit, where $I_1$ corresponds to the Euclidean global BTZ solution with temperature $\beta$. This solution topologically closes the asymptotic time circle $S^1_\beta$ in the bulk. To find $I_n(\varphi)$ we will look for bulk solutions that keep the $\mathbb{Z}_n$ replica symmetry. While each sheet of $\mathcal{M}_n$ is continued to the bulk, the replicated line $A=[0,\varphi]$ is continued inside to a replicated surface $\mathcal{E}^n_A$, together forming a smooth geometry.
The authors of  \cite{Lewkowycz:2013nqa} showed how to analytically continue this solution in $n$ and that as $n\rightarrow 1$ the difference in \eqref{eq:renyi_grav} localize to the boundary of the replicated surface $\mathcal{E}_A=\lim_{n\rightarrow 1}\mathcal{E}^n_A$. Taking the limit carefully gives the RT formula $S_{vN} = \frac{\text{Length}(X)}{4G_N}$, where we denoted the boundary of the bulk surface by $X = \partial \mathcal{E}_A$. The equation of motion for the metric constrain $X$ to be an extremal line (a geodesic). As $X$ is the boundary of the replicated surface $\mathcal{E}_A$, it is also constrained to be homologous to the boundary replicated line $A$. This is known as the `homology constraint' of RT surfaces \cite{Haehl:2014zoa,Rangamani:2016dms}.

\begin{figure}
	\centering
	\includegraphics[width=.6\linewidth]{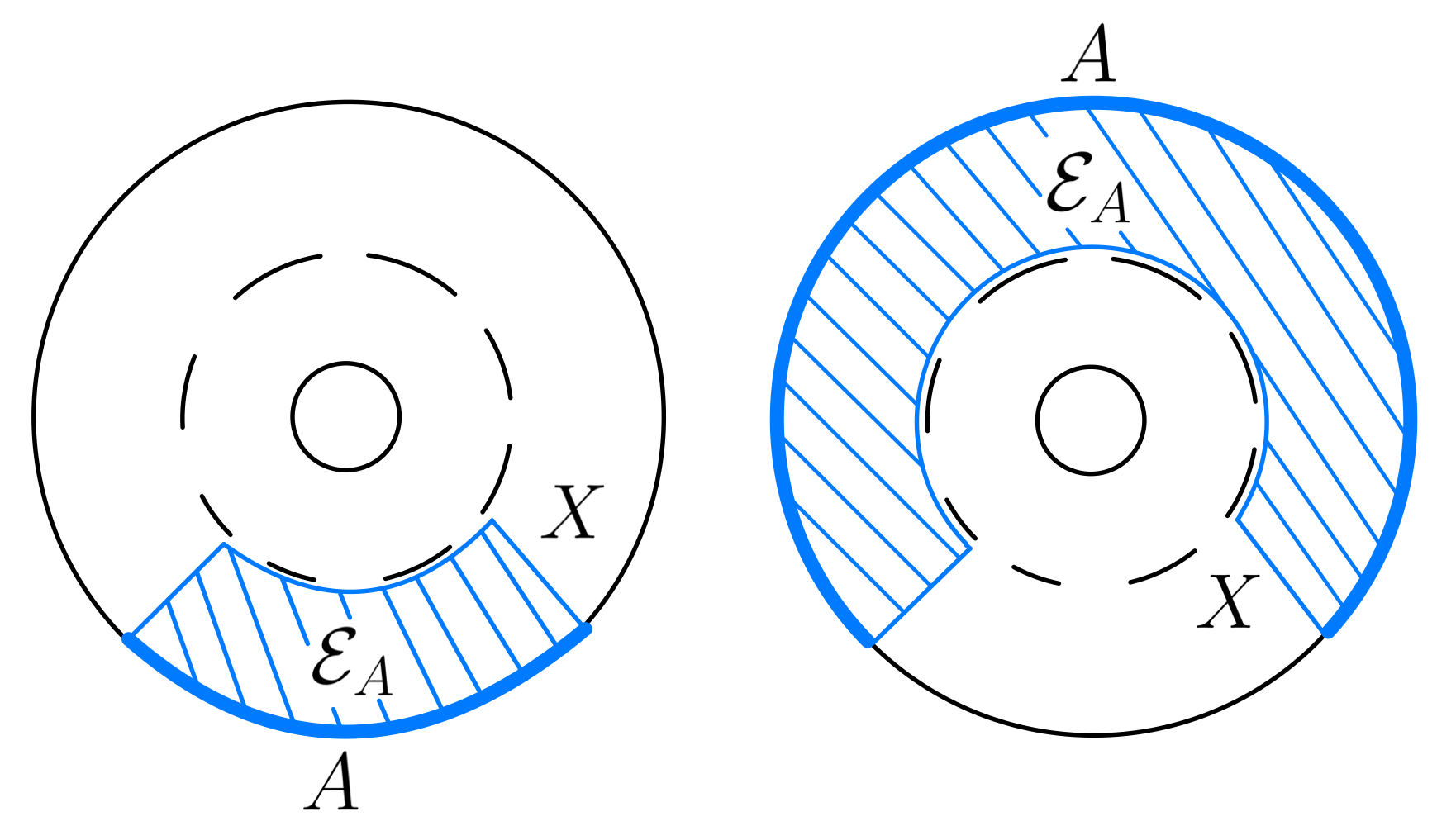}
	\caption{The spatial slice of BTZ black hole. The outer circle is the $\tau=0$ boundary spatial slice. The inner circle is $\tau=\pm \frac{\beta}{2}$ that we identify. The middle dashed circle is the (Euclidean) black hole horizon. The thick blue lines are two different boundary replicated lines $A$. The surface $\mathcal{E}_A$ is the extension of the replicated line inside the bulk. The blue line $X$ is its boundary, and the would-be extremal RT surface.
	\label{fig:thermal_RT}}
\end{figure}

In our case, we need to find Euclidean BTZ space-like geodesic $X$ that ends asymptotically at $\partial A$ and are homologous to $A$. For every $\varphi$ and high enough temperatures $\beta\ll 1$ there is only one solution.\footnote{For large enough $\varphi\sim2\pi$ (or low enough temperatures) there's a dominant disconnected solution with the same homology class \cite{Azeyanagi:2007bj} (for the relation to the Araki-Lieb inequality see also \cite{Headrick:2007km,Hubeny:2013gta}). \label{foot:disc_sol}} The geodesics are drawn schematically in figure \ref{fig:thermal_RT}. Calculating the length gives  \cite{Rangamani:2016dms}
\begin{equation}\label{eq:thermal_ee_holo}
	S_{vN}\left(\overline{\rho}(\varphi)\right) = \frac{c}{3} \log \left( \frac{\beta}{\pi a} \sinh\left( \frac{\varphi}{2\beta}\right)\right),
\end{equation}
$a$ being a bulk UV regulator. At the limit $\beta\ll \varphi$ we get back \eqref{eq:thermal_ee} as drawn schematically in figure \ref{fig:page_curv}. The linear behavior is explained geometrically by the geodesic roughly `hug' the black hole horizon for an arc of an angle $\varphi$. At $\varphi=2\pi$ the RT surface is the black hole horizon itself, which leads to the Bekenstein-Hawking formula for the thermodynamic entropy $S_{th}(\beta)=\frac{A_{BH}}{4G_N}$. In terms of the gravitational replica trick, the $\varphi=2\pi$ calculation is exactly that of Gibbons and Hawking  \cite{Gibbons:1976ue}.

\section{Typical pure state}\label{sec:typical}
\subsection{QFT side}
In the previous section we studied the von Neumann entropy of the averaged density matrix. Here we would like instead to find the ensemble average of the von Neumann entropy
\begin{equation}
	\overline{S_{vN}(\rho(\varphi))}= -\overline{\text{Tr}_A\left(\rho(\varphi)\log \rho(\varphi)\right)}.
\end{equation}
For brevity, we will denote this quantity (and its \myrenyi analogs) by $\overline{S_{vN}(\varphi)}$.
On cases where the ensemble represents typical states (see above), $\overline{S_{vN}(\varphi)}$ describes the von Neumann entropy of a typical pure state with energy $E\sim 1/\beta$. We will calculate the averaged entropy as the limit of the averaged \myrenyi entropy $\overline{S_{vN}(\varphi)}=\lim_{n\rightarrow 1} \overline{S_n(\varphi)}$.

For an ensemble state $\ket{\psi_\alpha}$, the \myrenyi entropy can be written as
\begin{equation}
	S_n(\rho_\alpha(\varphi)) = \frac{1}{1-n}\left(\log Z_n^\alpha - n \log Z_1^\alpha \right).
\end{equation}
where $Z_1^\alpha=\text{Tr}\rho_\alpha$ and 
$Z_n^\alpha = \text{Tr}_A \left(\text{Tr}_{A^c}\rho_\alpha\right)^n$. 
Taking the average gives
\begin{align} 
	\overline{S_n(\varphi)} & = \frac{1}{1-n}\left(\overline{\log Z_n} - n \overline{\log Z_1}\right)\label{eq:av_renyi_exact}\\
	& \approx \frac{1}{1-n}\left(\log \overline{Z_n}-\log \overline{Z^n_1}\right).\label{eq:av_renyi}
\end{align}
In the second line we took a simplifying assumption, that the quantities self-average well enough so we can take the average inside the log. This assumption is for brevity only. In appendix \ref{app:another_replica} we show how to use a second replica trick to calculate \eqref{eq:av_renyi_exact} exactly. For holographic theories the approximation is shown to be exact at leading (classical) order in $1/N$.\footnote{I thank Ohad Mamroud for emphasizing this point.} 

Following our expression for the thermal case \eqref{eq:thermal_renyi}, we can now write \eqref{eq:av_renyi} in terms of a path integral
\begin{equation}\label{eq:av_renyi_path}
	\overline{S_n(\varphi)} \approx 
	\frac{1}{1-n} \log \frac{
		\int_{\mathcal{M}_n} D\phi \ e^{-S[\phi]} 
		\ \delta\left(\left\{\mathcal{O}_i\left(\tau=\pm\frac{\beta}{2}\right)\right\}\right)
		}{
		\int_{\left(\mathcal{M}_1\right)^n} D\phi \ e^{-S[\phi]} 
		\ \delta\left(\left\{\mathcal{O}_i\left(\tau=\pm\frac{\beta}{2}\right)\right\}\right)
	}.
\end{equation}
By $\delta(\{\mathcal{O}_i\})=\prod_{i=1}^{n-1} \delta(\mathcal{O}_i-\mathcal{O}_{i+1})$ we mean a projector to states with the same $\mathcal{O}$ eigenvalues on all the $n$ copies. Apart from the identification at $\tau=\pm\frac{\beta}{2}$, we got the same path integrals as in the thermal case \eqref{eq:thermal_renyi}. 
Through the identification, the expression now depends on the ensemble basis we chose (unlike the thermal case). Specifically in the local field basis $\mathcal{O}=\{\hat\phi(\theta)\}$ the delta-function take a simpler form, of identifying the field's value between all the replicas on $\tau=\frac{\beta}{2}$.\footnote{For gauge theories (and orbifolds) a similar path integral can be written. Locally the identification has the same form as the scalar. Globally one also need to integrate over the $n$ different $S^1_\beta$ holonomies from each interval.}
As the ensemble average is over pure states we expect $\overline{S_{vN}(\varphi)}=\overline{S_{vN}(2\pi-\varphi)}$. In fact this property is immediate from the topology of the manifold on \eqref{eq:av_renyi_path} (unlike the thermal topology of \eqref{eq:thermal_renyi}).

\begin{figure}
\centering
	\begin{minipage}[c]{0.06\textwidth}
		$\text{Tr}\rho_\alpha^n=$
	\end{minipage}
	\begin{minipage}[c]{0.5\textwidth}
		\includegraphics[width=\linewidth]{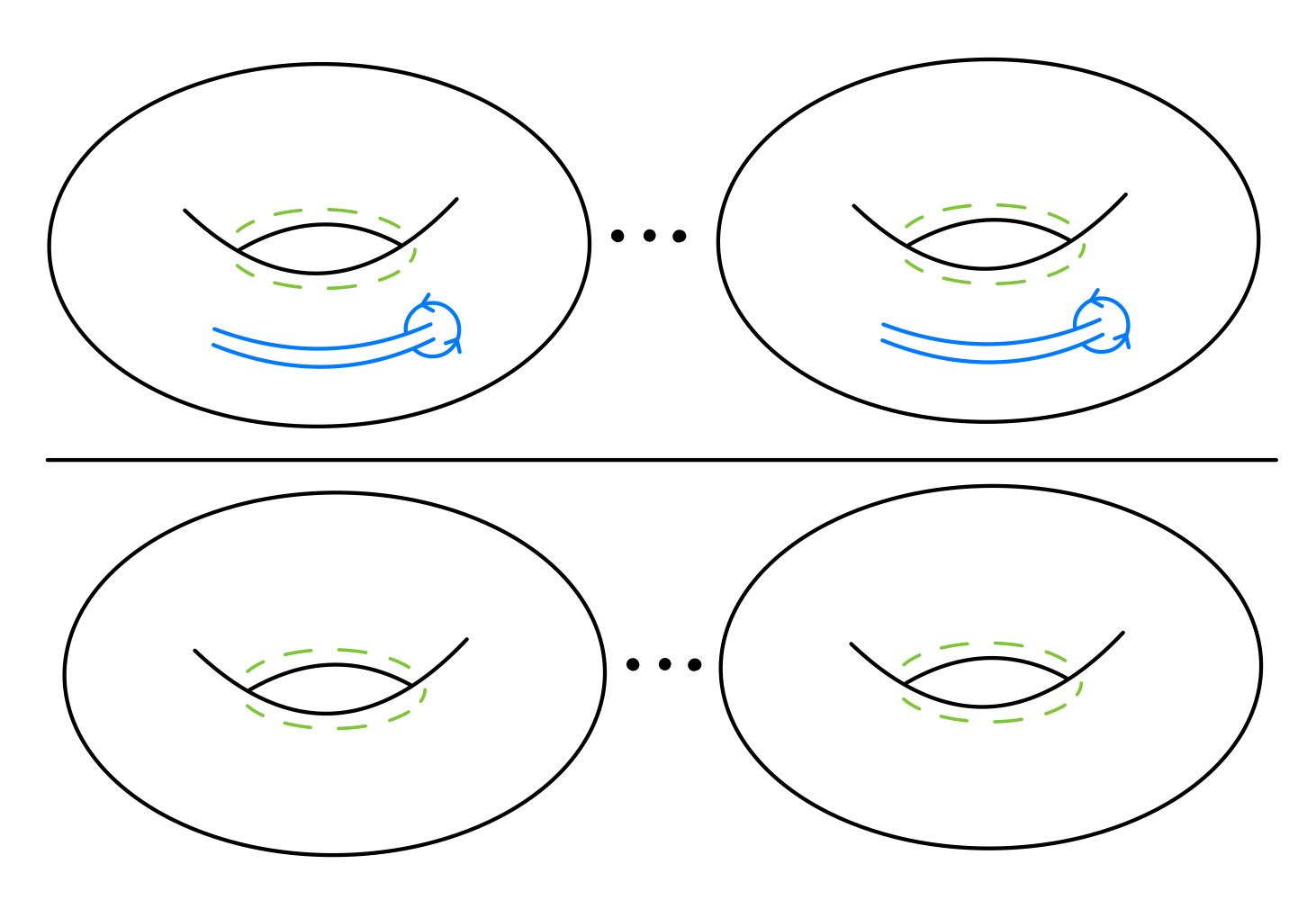},
	\end{minipage}
	\caption{The replica trick for pure ensemble state $\ket{\psi_\alpha}$. The green dashed circle represent the same boundary condition for all the replicas on the inner circle. The topology of the resulted manifold ensures $S_{vN}(\rho_\alpha(\varphi))=S_{vN}(\rho_\alpha(\varphi))$.
	\label{fig:pure_replica}}
\end{figure}

The reader may worry that the identification between the replicas in \eqref{eq:av_renyi_path} makes the expression singular or even ill-defined. As the singularities are coming from a local theory it is enough to consider the denominator integral in \eqref{eq:av_renyi_path}, which includes no replicated cut. In appendix \ref{app:ident} we study its behavior for the free scalar. The singularities from the identifications seem to be basis-dependent. These UV singularities can be regulated for example by a cutoff or the zeta-function regularization. In terms of a UV cutoff, the path integral can be regulated by the normal local counter-terms on the full manifold (as in the thermal case), together with new local counter-terms on the $n-1$ identified circles. Both because of the basis-dependence and due to the geometries we will consider, it is not clear what is the bulk interpretation of these divergences/counter-terms (see below).

What is the expected behavior for $\overline{S_{vN}(\varphi)}$?  For $\beta\gg 1$ the ensemble reduce to the ground-state, and $\overline{S_{vN}(\varphi)}$ takes the universal ground-state result \eqref{eq:vac_ee}. When the entanglement region is very small $\varphi\ll1,\beta$ we can use the twist operator OPE and recover the ground-state result \eqref{eq:vac_ee}
\begin{equation}
	\overline{S_{vN}(\varphi)} \approx \frac{c}{3}\log\left(\frac{\varphi}{a}\right).
\end{equation}
We are interested in the high temperatures limit $\beta\ll 1$. In the strict limit $\beta\ll a$ (the lattice regulator) the ensemble states are exactly the field states $\ket{\psi_\alpha}=\ket{\phi_\alpha}$. 
Upon lattice regularization, these states are products of lattice-points position states giving $\overline{S_{vN}(\varphi)}=0$.
At intermediate temperatures $a \ll \beta \ll \varphi$ we expect a thermal behavior \eqref{eq:thermal_ee}. The intuition is that we trace over a big reservoir of a high-temperature pure state.\footnote{This property was termed `canonical typicality' \cite{Goldstein:2005aib}. Importantly, the microcanonical ensemble would differ from the canonical one for the \myrenyi entropies but will agree on the von Neumann entropy \cite{Lashkari:2016vgj,He:2017vyf} (and also found holographically \cite{Dong:2016fnf,Dong:2018lsk}). It is therefore essential that our ensemble does not consist of energy eigenstates. \label{foot:typicality}}
From the symmetry around $\varphi=\pi$, we expect a Page-like behavior for high-enough temperatures $a\ll \beta \ll 1$ (see figure \ref{fig:page_curv}).
In the next section, we find $\overline{S_{vN}(\varphi)}$ for holographic large-$N$ theories using semiclassical gravity. Based on either subsystem ETH \cite{Dymarsky:2016ntg} at $\beta \ll 1$ or large central charge expansion \cite{He:2017vyf} we expect the pure state semiclassical result to exactly match the thermal holographic result \eqref{eq:thermal_ee_holo} for $\varphi<\pi$. Below we will argue this is actually the case using the holographic dictionary.

\begin{figure}
\centering
	\includegraphics[width=\linewidth]{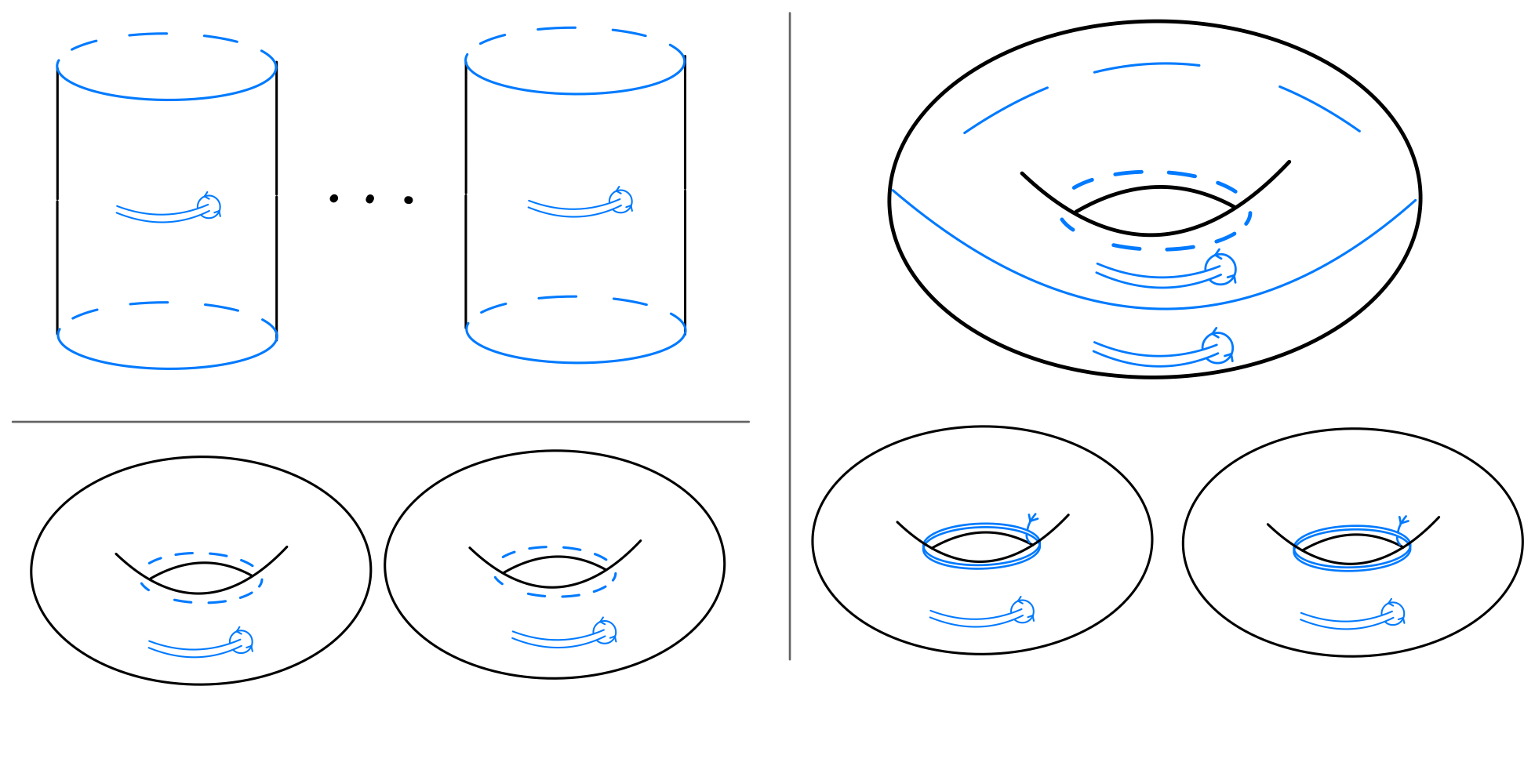},
	\caption{Top left: $n$ cylinders with a replicated line. All the cylinder's boundaries are identified (dashed blue line). Bottom left: Each cylinder is understood as a torus. $\tau=\pm \frac{\beta}{2}$ is identified between the $n$ tori. Top right: stacking the $n=2$ cylinders into a bigger torus with temporal periodicity $2\beta$. the circles at times $\tau=0,\beta$ are identified. Bottom right: the same topology can be drawn as $n=2$ tori with another replicated circle at $\tau=\pm\frac{\beta}{2}$.
	\label{fig:topologies}}
\end{figure}

Before turning to the gravity side, we need to describe the topology of the manifolds in \eqref{eq:av_renyi_path}. 
We claim that several different smooth geometries exist, all giving the same path integral \eqref{eq:av_renyi_path}.
As our main example we take the field states basis. The basic topology one gets from the canonical formulation is that of ($S^{d-1}$ times) $n$ intervals $\left[-\frac{\beta}{2},\frac{\beta}{2}\right]^n$ with field identifications between all their $2n$ boundaries (top left in figure \ref{fig:topologies}). Specifically the field derivative might jump when crossing $\tau=\pm\frac{\beta}{2}$. 
For the thermal case $n=1$, we argued above that we can assume the smoothness of the fields turning the interval topology to a circle, with equivalent results.
Correspondingly, we can make different smoothness assumptions on the $n$ intervals changing the topology, with the same results. One option is to assume the smoothness of the fields when crossing at the same time interval, which results in the topology of $n$ tori (bottom left in figure \ref{fig:topologies}). Between the $n$ tori we still need to identify the fields at $\tau=\pm\frac{\beta}{2}$, which still allows a derivative jump when crossing to another torus.
A second option we will consider is to stack the intervals into one long circle of length $n\beta$. We then take the field to be smooth along it, which results in a topology of a single long torus (top right in figure \ref{fig:topologies}). Along the torus we still identify the spatial-slices of $\tau=0,\beta,...,(n-1)\beta$, but not their derivative. We can also think of this topology as $n$-sheeted torus (with time circle of length $\beta$) with a replicated circle at $\tau=\pm \frac{\beta}{2}$ (bottom right in figure \ref{fig:topologies}). We conclude that the two smoothness assumptions give rise to two different topologies. Specifically note that the first has $n$ connected components (for $\varphi=0$), while the second is connected (see figure \ref{fig:topologies}). The path integral on one will be equal to the second, and both to the original cylinder geometry of \eqref{eq:av_renyi_path}.

One can make the same smoothness assumptions also for the numerator in \eqref{eq:av_renyi_path}, which will add a replicated line along $A=[0,\varphi]$ (and $\tau=0$) to the same geometries. In fact, starting from one of the two geometries at some $\varphi=\varphi_0$, and continuously deforming the cut to $\varphi=2\pi-\varphi_0$ will result in the second geometry of $\varphi=\varphi_0$.
The two compact topologies described are the only ones that respect the $\mathbb{Z}_n$ replica symmetry. There are more complicated topologies (correspond to other smoothness assumptions) that break the symmetry, which we won't describe here. Following the thermal $n=1$ case, we argue all the different topologies are equivalent in the path integral, and equivalent to the original description where no smoothness assumptions were assumed.\footnote{In principle, classical (field theory) saddles are allowed to be non-smooth at the identification, or smooth only in several of the topologies.} 
We stress that we don't mean one needs to sum over all these topologies to get the right result in \eqref{eq:av_renyi_path}. We are claiming that the path integrals are equal, and the path integral on either one of the geometries (together with the remaining identifications) is already the right result. We show it explicitly at $n=2$ (two replicas) for the free theory in appendix \ref{app:ident}.

\subsection{Gravity side}
We now turn to calculate the averaged entropy $\overline{S_{vN}(\varphi)}$ using holography. Following the thermal calculation reviewed in section \ref{sec:termal_GR}, we will do it by analytically continuing the gravitational calculation of the averaged \myrenyi entropies \eqref{eq:av_renyi_path}. The first step is to write the path integrals in \eqref{eq:av_renyi_path} as gravitational path integrals. At the CFT we found several equivalent compact topologies corresponding to different smoothness assumptions. To calculate the dual gravitational path integral we will assume that the main contribution to the path integral comes from smooth geometries that satisfy the boundary conditions. In this way, we get different saddles from each boundary topology. But unlike the QFT side, the saddles from each boundary topology won't be equivalent to each other anymore and will give different results. We conclude that on the gravity side one needs to sum over saddles from all the allowed boundary topologies. As a sanity check, note that in this way the $\varphi \mapsto 2\pi-\varphi$ symmetry of \eqref{eq:av_renyi_path} is preserved also in the bulk (see below).

Let's be more concrete. Every smoothness assumption on the QFT side leaves a path integral over a smooth compact geometry together with extra identifications (the dashed blue lines in figure \ref{fig:topologies}). We are looking for smooth gravitational saddles with that compact geometry as its asymptotic boundary. But we also need to map the remaining identifications of the fields to the gravity side. The identification gives further boundary conditions on the background SUGRA  fields (or any other low energy bulk description) solution. To see it, notice that since all the CFT fields are identified between the replicas (at $\tau=\frac{\beta}{2}$), so is the CFT stress tensor $T_{\theta,\theta}(\theta,\tau=\frac{\beta}{2})$. As a result, the boundary mode of the bulk metric $g_{\theta,\theta}(\theta,\tau=\frac{\beta}{2},z=0)$ is identified between the $n$ replicas. We expect similar boundary conditions for all the bulk's low-energy SUGRA fields \cite{Witten:2001ua}. Note that this is the boundary condition dual to the field-basis ensemble $\mathcal{O}=\{\phi(\theta)\}$. Other boundary conditions will depend on the bulk dual of the $\mathcal{O}_i$ in \eqref{eq:av_renyi_path}. We expect that for `local enough' operators, we would have a similar picture.\footnote{Specifically, we don't expect it to work for the microcanonical ensemble (see also footnote \ref{foot:typicality}). As the gluing in this case will include non-local boundary identifications, it is no longer clear smooth geometries dominate the partition function.}
The bulk geometries we will consider below all preserve the replica symmetry in the bulk (see appendix \ref{app:another_replica}). As a result, as long as we are allowed to use the semiclassical approximation for \eqref{eq:av_renyi_path}, these saddles will satisfy the necessary boundary conditions between the replicas.
Although replica-breaking solutions exist, far enough from $\varphi=\pi$ the replica-symmetric solutions are the dominant ones (see appendix \ref{app:another_replica}). 
The smoothness assumption thus sets the geometry and will give the entropy at tree level. The extra (ensemble-dependent) bulk field identifications will only affect the subleading 1-loop order. We will comment on that at the end of the section.
In any case, we conclude that to leading order in $G_N$, our results are robust to the specifics of the ensemble.

\begin{figure}
	\centering
	\includegraphics[width=.6\linewidth]{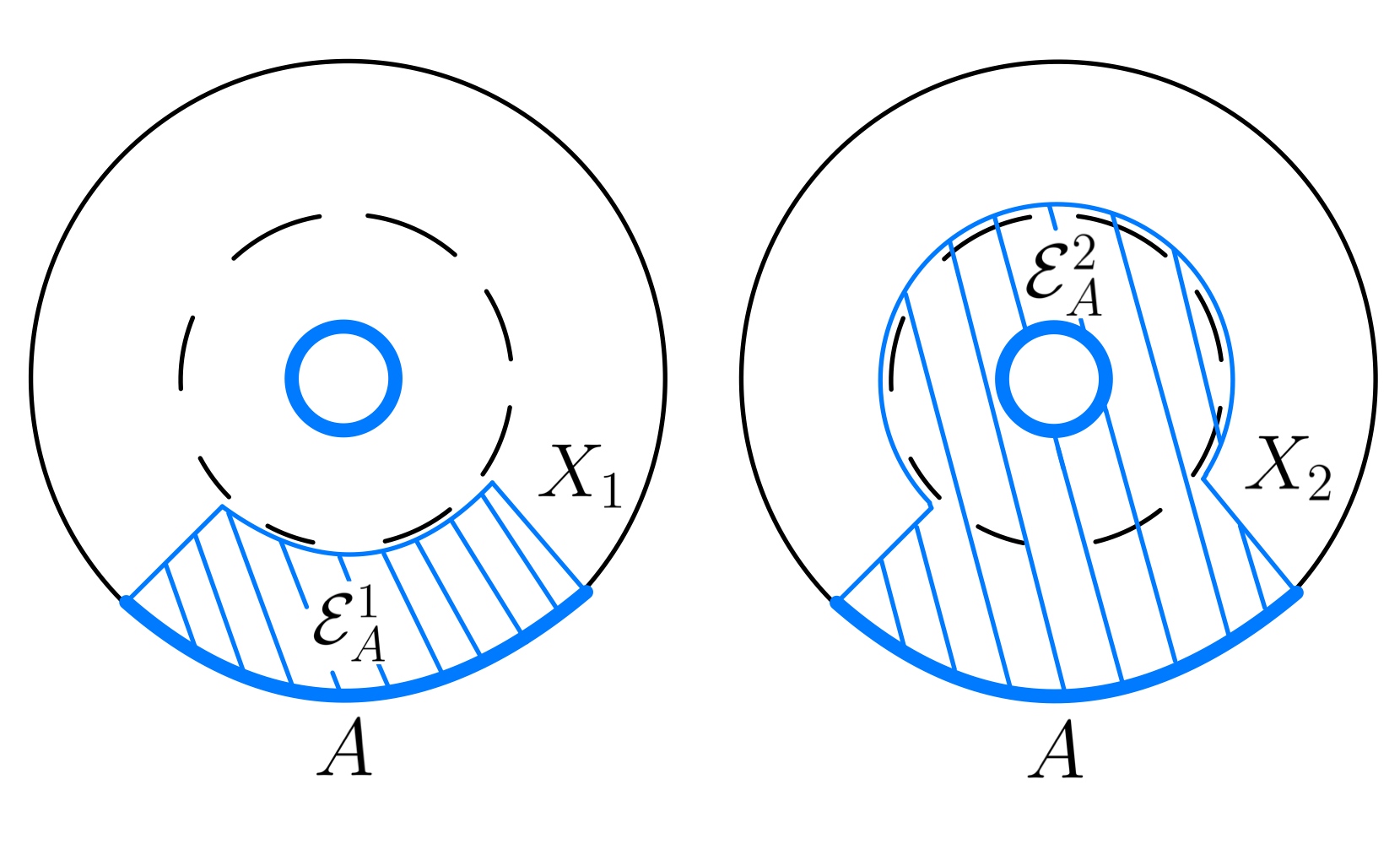}
	\caption{Here the inner circle (thick blue) $\tau=\pm \frac{\beta}{2}$ is identified between all the replicas. On left the same bulk replicated surface $\mathcal{E}^1_A$ found for the thermal case figure \ref{fig:thermal_RT}. On the right a second replicated surface $\mathcal{E}^2_A$ of a different topology. This solution is allowed only due to the inner circle identification.
	\label{fig:pure_RT}}
\end{figure}

Looking back at \eqref{eq:av_renyi_path}, there are two candidates for dominant bulk saddles at $\beta\ll 1$ corresponding to the two replica-symmetric boundary topologies.
The first boundary topology is that of $n$ separated tori. This is the same boundary condition for the metric as in the thermal calculation \eqref{eq:renyi_grav}. The solution includes an extension of the replicated line to a bulk replicated surface $\mathcal{E}^1_A$, see figure \ref{fig:thermal_RT}. Whenever this saddle dominates, the averaged entropy (after taking the $n\rightarrow 1$ limit) is just the thermal entropy \eqref{eq:thermal_ee_holo}.

\begin{figure}
	\centering
	\includegraphics[width=.6\linewidth]{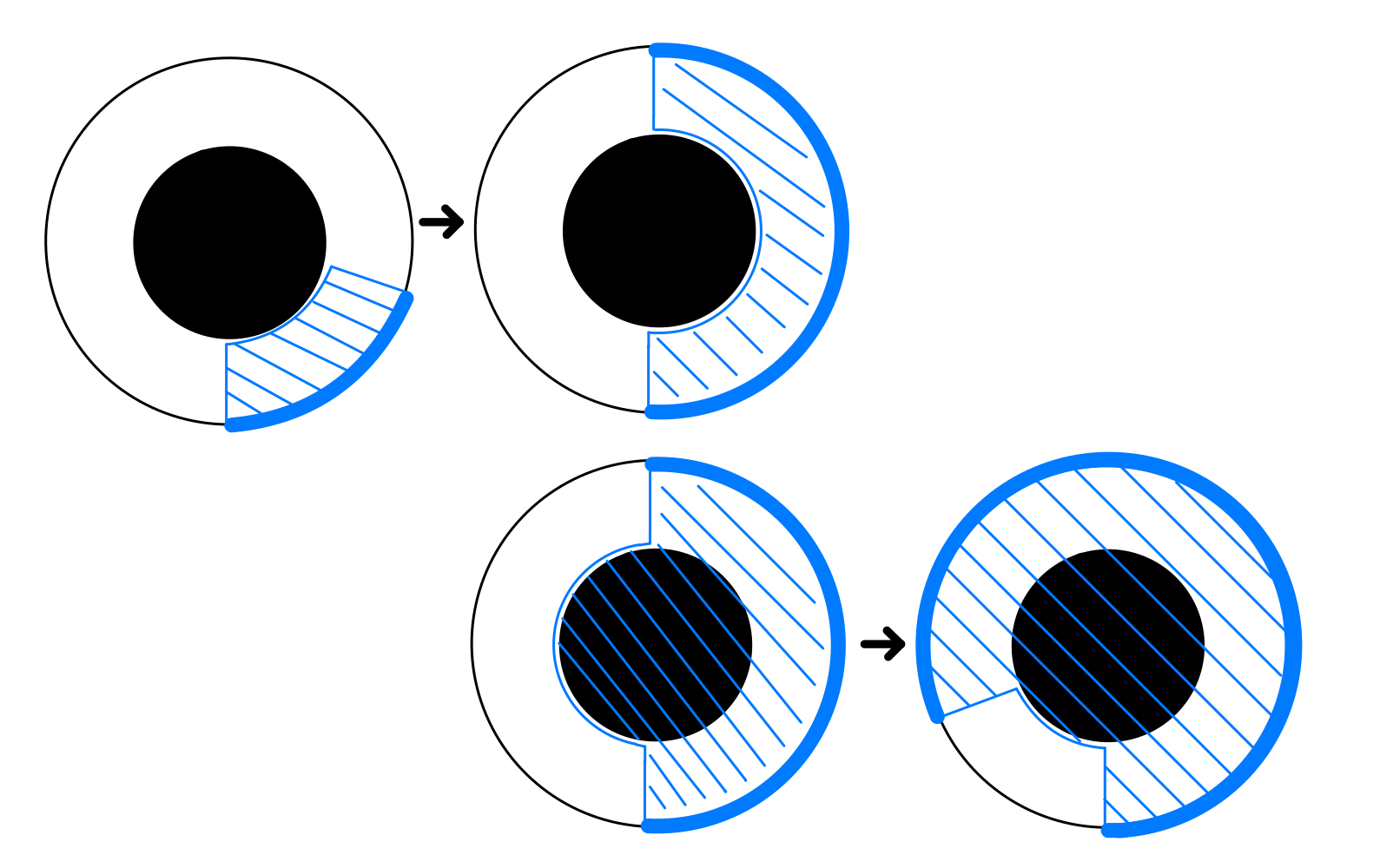}
	\caption{The phase transition of the RT surface as a function of $\varphi$. The outer circle is the boundary spatial slice $\tau=0$. The black solid circle is the BTZ black hole (it's boundary is the horizon). The upper row include the first `thermal' RT surface (thin blue line) changes as a function of $\varphi$. The middle column describe the phase transition at $\varphi=\pi$ to the new solution on the bottom row. Together the two solutions give a Page-like behavior for the holographic $\overline{S_{vN}(\varphi)}$ (see figure \ref{fig:page_curv}).
	\label{fig:phase_transition}}
\end{figure}

The second boundary topology is that of one long torus, with $n$ replica cuts along it.
Equivalently it can be described as $n$ separated tori, connected by both the replicated line $A$ and another replicated circle at $\tau=\frac{\beta}{2}$. 
Now the gravitational solution $I_n(\varphi)$ will include a replicated surface $\mathcal{E}^2_A$ which extend both replicated lines together. In other words its asymptotic boundary is both $A$ and the $\tau=\pm \frac{\beta}{2}$ circle. Following the same logic of  \cite{Lewkowycz:2013nqa}, at the limit $n\rightarrow 1$ the contribution of this saddle would be an RT line $X_2$ which is homologous to the sum of region $A$ and the identification circle (see figure \ref{fig:pure_RT}). Equivalently in terms of the bulk spatial slice, the line is homologous to the complement $A^c = [\varphi, 2\pi]$. Therefore whenever this saddle dominates, it contributes exactly thermal entropy of the complement $A^c$. Taking only the dominant saddle at leading order\footnote{As we mention in footnote \ref{foot:disc_sol} both topologies allow another solution with a disconnected RT line covering the horizon. For the second (non-trivial) boundary topology this is explicitly a replica island covering the horizon. As these solutions are always sub-dominant we omit them in \eqref{eq:pure_entropy_holo}.} gives together\footnote{For states similar to our \eqref{eq:ensemble_state}, \cite{Asplund:2014coa} found the same result using a large central charge expansion in $\text{CFT}_2$.}
\begin{equation}\label{eq:pure_entropy_holo}
	\overline{S_{vN}(\varphi)} = \text{min}\left\{\frac{c}{3} \log \left( \frac{\beta}{\pi a} \sinh\left( \frac{\varphi}{2\beta}\right)\right),
	\frac{c}{3} \log \left( \frac{\beta}{\pi a} \sinh\left( \frac{2\pi-\varphi}{2\beta}\right)\right) \right\}.
\end{equation}
As the thermal result is monotonic in $\varphi$ it is dominant only for $\varphi<\pi$. For $\varphi>\pi$ the second non-trivial topology dominates. In the high-temperature limit $\beta \ll \varphi$ we have 
\begin{equation}
	\overline{S_{vN}(\varphi)} \approx \frac{c}{6\beta} \min\{\varphi,2\pi-\varphi\},
\end{equation}
as drawn schematically in figure \ref{fig:page_curv}. This is the holographic dual of the calculation made by Page \cite{Page:1993df}. We can further draw this phase transition in terms of $\mathcal{E}_A$. As can be seen in figure \ref{fig:phase_transition}, before the phase transition $\mathcal{E}_A$ covers only the exterior of the black hole, but `swallows' the black hole after the transition. Finally, we note that at the low-temperature phase the geometry is of thermal AdS. In this case there's no phase transition and the disconnected replica  solution $\mathcal{E}^1_A$ is always dominant, leading to the vacuum result \eqref{eq:vac_ee}. In other words at this order in $1/G_N$ the extra identification is obsolete.

We learn that the `purity' of the boundary geometry \eqref{eq:av_renyi_path} takes place in the bulk by the fact that any solution for $\varphi$ can be transformed to a solution of $2\pi-\varphi$. This way the gravitational path integral ensures the purity of the result $\overline{S_{vN}(\varphi)}=\overline{S_{vN}(2\pi-\varphi)}$. Consider now a general (compact) spatial manifold $\mathcal{M}^{d-1}$ and a general entanglement region $A\subset \mathcal{M}^{d-1}$. Taking $n\rightarrow 1$ for the generalization of \eqref{eq:av_renyi_path} calls for a refinement of the known RT formula for typical pure states of energy $E \sim 1/\beta$:
\begin{equation}\label{eq:typical_RT_proposal}
	\overline{S_{vN}(A)}_\beta = \frac{\text{Area}(X^*)}{4G_N} + O(G_N^0).
\end{equation}
Just like in the RT formula, $X^*$ is a minimal co-dimension $2$ manifold in the bulk solution (with boundary $\mathcal{M}^{d-1}\times S^1_{\beta}$) which asymptote to $\partial A$. For the typical state, we further allow $X^*$ to be homologous to either $A$ or its spatial complement $A^c$. In other words, for typical pure states the `homology constraint' should be relaxed.

Following  \cite{Faulkner:2013ana}, we can study the next order in $1/G_N$. To that end we consider the 1-loop contribution from bulk fields around the dominant gravitational saddle. Ignoring the effect of the identification on the fields, the path integral prepare a thermofield double state on the bulk global spatial slice, or a bulk thermal density matrix on the slice that ends at the horizon. But as we discussed above the CFT identification on \eqref{eq:av_renyi_path} is dual to the identification of the field's asymptotic value at $\tau=\pm\frac{\beta}{2}$ between the $n$ bulk replicas. Each asymptotic value prepares a state at $\tau=-\frac{\beta}{2}$. The bulk state at the slice $\tau=0$ is then the bulk EFT evolution of this state by $e^{-\frac{\beta}{2} H_{bulk}}$. We learn that each CFT ensemble corresponds to some bulk effective ensemble of `almost-thermal' density matrices. Following  \cite{Faulkner:2013ana}, we propose that the next order is\footnote{To write the second term we actually need the original exact expression \eqref{eq:av_renyi_exact}.}
\begin{equation}\label{eq:1loop_proposal}
	\overline{S_{vN}(A)}_\beta = \frac{\text{Area}(X^*)}{4G_N} + \overline{S^\text{bulk}_{vN}(\mathcal{E}^*_A)}_\beta + O(G_N^{-1}).
\end{equation}
The ensemble average on the RHS is in the bulk EFT using the bulk dual EFT ensemble. Note that from the prescription for $X^*$, $\mathcal{E}^*_A = \mathcal{E}^*_{A^c}$ which ensures also the purity of the second term in \eqref{eq:1loop_proposal}.

We note that in the full string theory we will need to further identify all the heavier vertex operators insertions between the identification circles. The exact identification will depend on the exact string duals of the operators $\mathcal{O}$. As we identify the full string state between the asymptotic temporal circles, the string theory path integral also allows for new types of non-dynamical strings. 
Close strings can stretch between (asymptotic) identified circles of the same geometry. Strings can also end in one identified circle and then continue on a circle of another disconnected geometry.\footnote{These strings are dual to the gauge-invariant holonomies measured along a line between two identified points.} 
These are not orbifolds or D-branes, but a complete identification of spacetime (asymptotic) points. This identification might be related to the localized singularities we find in the QFT calculation of \eqref{eq:av_renyi_path} and described in appendix \ref{app:ident}. Further research is needed to determine the consistency of string theories with such identifications, and we won't explore them further here.\footnote{I thank Ofer Aharony for stressing these issues.}

\section{Discussion}\label{sec:discussion}
Following the idea of `entanglement wedge reconstruction' \cite{Dong:2016eik}, we can interpret the phase transition of $\overline{S_{vN}(\varphi)}$ in terms of bulk reconstruction from the CFT. As a function of $\varphi$ the entanglement wedge, the domain of dependence of $\mathcal{E}_A^*$, include the black-hole interior starting from $\varphi>\pi$ (figure \ref{fig:phase_transition}). We learn that it requires operators from half the boundary area to locally describe the black hole interior (to the extent such a description is possible). Note that this is possible only for a pure state and not for the thermal density matrix. 
There are different ideas for explicit boundary-to-interior mappings ~\cite{Papadodimas:2013jku,Penington:2019kki}.
It would be interesting to see if such mappings are possible in the state ensemble context (or they are too state-dependent to survive the average).

The analysis above was agnostic to the correct quantum-gravitational description of the black hole interior. It would be interesting however to find a way to connect this work to older discussions about typical black-hole microstates in holography \cite{Marolf:2013dba,deBoer:2019kyr,Bousso:2013wia}, as well as the new discussions in the context of evaporation \cite{Almheiri:2019psf,Penington:2019kki,Almheiri:2019qdq}.

The HRT prescription extends the RT formula for general density matrices including arbitrary time dependence \cite{Hubeny:2007xt,Dong:2016hjy}. Recently the idea of replica wormholes was similarly generalized to real-time gravitational path integrals \cite{Goto:2020wnk,Colin-Ellerin:2020mva,Colin-Ellerin:2021jev}. Using these methods, it would be interesting to find a Lorentzian version of our analysis, perhaps to study the averaged effect of infalling matter into a typical black hole microstate.

Finally, we can take our assumption about the leading gravitational saddles to $\overline{Z_n}$ as a property of holographic CFTs. In the CFT language at large $N$ it gives the equality between the $\overline{Z_n}$ path integral and a sum of CFT path integrals over different partitions of the replicas (without any further identifications). This equality was recently found directly in the CFT (in a slightly different context) by assuming ergodicity properties~\cite{Pollack:2020gfa,Liu:2020jsv,Sasieta:2021pzj,Freivogel:2021ivu}. It would be illuminating to study further the state-averaged gravitational path integral as a probe of the dual CFT statistical properties~\cite{Belin:2020hea}.

\section*{Acknowledgements}

I would like to thank Micha Berkooz, Anatoly Dymarsky, Rohit Kalloor, Adar Sharon, Ohad Mamroud, Tal Sheaffer, and Masataka Watanabe for useful discussions, and especially Ofer Aharony for many helpful discussions and guidance. I also thank Ofer Aharony, Micha Berkooz, Anatoly Dymarsky, Ohad Mamroud, Mukund Rangamani, and Adar Sharon for their comments on a draft of the manuscript.
The work was supported in part by an Israel Science Foundation center for excellence grant (grant number 2289/18), by grant no.2018068 from the United States-Israel Binational Science Foundation (BSF), and by the Minerva foundation with funding from the Federal German Ministry for Education and Research.

\appendix

\section{Double replica trick} \label{app:another_replica}
The averaged \myrenyi entropy is given by \eqref{eq:av_renyi_exact} $\overline{S_n(\varphi)} = \frac{1}{1-n}\left(\overline{\log Z_n} - n \overline{\log Z_1}\right)$.
In order to calculate the first term, we can use a second replica trick
\begin{equation}\label{eq:sec_rep}
	\overline{\log Z_n} = \lim_{m\rightarrow 1} \frac{1}{1-m} \log \overline{Z_n^m}.
\end{equation}
$\overline{Z_n^m}$ is a path integral over $m$ copies of the $n$-sheeted torus $\mathcal{M}_n$, with all the $n\cdot m$ $\tau=\pm\frac{\beta}{2}$ circles identified. 

What is the dominant bulk saddle of $\overline{Z_n^m}$? Above we assumed for $m=1$ the dominating solution is invariant under the replica symmetry $\mathbb{Z}_n$. We argue more generally that for any $m$, the dominant bulk saddle is $m$ times a replica-symmetric solution that fills $\mathcal{M}_n$. These are the solutions invariant under the full $S_m\times \mathbb{Z}_n$ symmetry of $\overline{Z_n^m}$. Notice that these solutions still respect the identification boundary conditions between the $m$ copies as it is symmetric under permutations $S_m$.

Start at $\varphi=0$, where $\overline{Z_n^m}$ has an $S_{m\cdot n}$ permutation symmetry. On general grounds the free energy on $S^{d-1}\times S^1_{\beta}$ for $\beta \ll 1$ is $I(\beta) \equiv -\log Z(\beta) = -\frac{a_d}{G_N \beta^{d-1}}$ for some constant $a_d$ \cite{Aharony:2019vgs}. The solutions at $\beta\ll 1$ correspond to different partitions of the $n\cdot m$ intervals into tori. The fully disconnected solution include $n\cdot m$ seperated tori with temporal length $\beta$ each, and the free energy $n\cdot m \cdot I(\beta) = -n\cdot m \cdot \frac{a_d}{G_N \beta^{d-1}}$. The fully connected solution is a torus with temporal length $n \cdot m \cdot \beta$, and free-energy $I(n m \beta) = -(n \cdot m)^{1-d} \cdot \frac{a_d}{G_N \beta^{d-1}}$. Any other partition $\{ n_i \}$ with $\sum_i n_i = n\cdot m$ has free energy $\sum_i I(n_i \beta) = \left(\sum_i n_i^{1-d}\right) \cdot \frac{a_d}{G_N \beta^{d-1}}$. Therefore at $\varphi=0$ the dominant solution is the fully disconnected one. At small enough $\varphi \ll 1$ we expect the dominant saddle to be a deformation of the fully disconnected saddle, which extends the replica cuts to the bulk (and thus connecting each of the $m$ copies). Without loss of generality we can take the minimal deformation on each of the separated $m$ copies. Inside each copy it is also reasonable to assume the solution respect the replica symmetry (as topologically it is already satisfied). For the case $m=1$ this is exactly the `thermal saddle' we considered above. Using the symmetry $\varphi \mapsto 2\pi-\varphi$ we also know the dominant solution at $2\pi -\varphi \ll 1$ is $m$ copies of the `fully connected saddle' (in terms of the thermal circle identification). A priori other saddles might dominate in the intermediate region between $0$ and $2\pi$. In practice, as they break the $S_m \times \mathbb{Z}_n$ symmetry we expect them to contribute only around $\varphi = \pi$.

Taking only the dominant solution gives $\overline{Z_n^m}=\overline{Z_n}^m$. Pluging inside \eqref{eq:sec_rep} brings $\overline{\log Z_n}=\log \overline{Z_n}$, as we approximated in \eqref{eq:av_renyi}. The same analysis shows that the variance in $\overline{S_{vN}(\varphi)}$ is non-perturbative in $G_N$. We note that if the bulk theory field content breaks the replica-symmetry, the breaking is perturbative in $G_N$.

\section{Details about the identification}\label{app:ident}
The goal of this section is to study the path integral on $N$ identified cylinders, called $\overline{Z_1^N}$ on \eqref{eq:av_renyi_path}. We will do the calculation for the free boson and study its divergences. We will also explicitly show it is equal to the path integral over the other two geometries described in figure \ref{fig:topologies}. We first practice on the harmonic oscillator, where we show the equivalence first for $N=1$ (the thermal case) and then for general $N$. We then follow to the free 2d boson.

\paragraph{Regularization} For all the calculations below we us the zeta-function regularization. Specifically we will be needed to regularize expressions like $\prod_{n=1}^{\infty}an^{\alpha}$. To that end we define
\begin{equation}
\begin{split}
A_{s} & =\sum_{n=1}^{\infty}\left(an^{\alpha}\right)^{s}\\
 & =a^{s}\zeta\left(-\alpha s\right).
\end{split}
\end{equation}
Using the zeta-function regularization
\begin{equation}
\begin{split}
\sum_{n=1}^{\infty}\log\left(an^{2}\right) & =\partial_s A_{s}\mid_{s=0}\\
 & =-\alpha\zeta'\left(0\right)+\zeta\left(0\right)\log\left(a\right)\\
 & =\log\left(\sqrt{\frac{\left(2\pi\right)^{\alpha}}{a}}\right),
\end{split}
\end{equation}
or
\begin{equation}
``\prod_{n=1}^{\infty}an^{\alpha}"=\sqrt{\frac{\left(2\pi\right)^{\alpha}}{a}}.\label{eq:zeta_eq}
\end{equation}

\subsection{Harmonic Oscillator}
We start by studying the Euclidean harmonic oscillator on $S^1_\beta$
\begin{equation}
	\begin{split}
	S\left[x\right]=\int_{0}^{\beta}d\tau\left(\frac{1}{2}m\dot{x}^{2}+\frac{1}{2}m\omega^{2}x^{2}\right).
	\end{split}
\end{equation}
Our main goal is to study the one-dimensional version of \eqref{eq:av_renyi} $\overline{Z_1^N}$ in the position basis, or simply the path integral over $N$ identified intervals. By the end of this section, we will show explicitly for $N=2$, that the euclidean path integral on the two identified intervals is equal to a path integral over two other geometries. The first is two identified circles, and the second is one doubly-long circle identified in the middle (see figure \ref{fig:free_calc}). We start by studying the thermal $N=1$, where we show explicitly that the path integral on a single identified interval is equal to the path integral over a circle.
We will use all these results for the free scalar in the next section.

\subsubsection{Thermal partition function}
	We start with the computation of the thermal partition function using the path integral formalism. Following the Hilbert-space definition, we need to find:
	\begin{equation}\label{eq:HO_Z}
	\begin{split}
		Z\left(\beta\right)& =\text{Tr}\left(e^{-\beta H}\right)\\
		&=\int_{-\infty}^{\infty}dx\bra{x}e^{-\beta H}\ket{x}\\
	\end{split}
	\end{equation}
	The Euclidean propagation $\bra{x}e^{-\beta H}\ket{x}$ can be written in the path integral formalism as
	\begin{equation}\label{eq:KE_def}
	\begin{split}
	\bra{x}e^{-\beta H}\ket{x}=\int Dx\left(\tau\right)\mid_{x\left(0\right)=x}^{x\left(\beta\right)=x}e^{-S\left[x\right]}.
	\end{split}
	\end{equation}
	We see that $Z(\beta)$ is strictly equal to the path integral over the identified interval $[0,\beta]$.

	We start by finding the Euclidean propagator $\bra{x}e^{-\beta H}\ket{x}$.
	The classical solution ($\ddot{x}=\omega^{2}x$) satisfying the
	boundary conditions is $x_{cl}\left(\tau\right)=x\left(\cosh\left(\omega \tau\right)+\left(1-\cosh\left(\omega\beta\right)\right)\frac{\sinh\left(\omega \tau\right)}{\sinh\left(\omega\beta\right)}\right)$. Changing variables in \eqref{eq:KE_def} by the shift $y\left(\tau\right)=x\left(\tau\right)-x_{cl}\left(\tau\right)$ gives
	\begin{equation}\label{eq:K_E_y}
	\begin{split}
	\bra{x}e^{-\beta H}\ket{x} & =e^{-S[x_{cl}]}\int Dy\left(\tau\right)\mid_{y\left(0\right)=0}^{y\left(\beta\right)=0}e^{-S\left[y\right]}\\
	 & =e^{-S[x_{cl}]}\bra{0}e^{-\beta H}\ket{0}
	\end{split}
	\end{equation}
	We diagonalize $S\left[y\right]$ by decomposing $y(\tau)$ into the orthonormal basis
	\begin{equation}\label{eq:y_basis}
	\begin{split}
	y\left(\tau\right) & =\sum_{n=1}^{\infty}a_{n}y_{n}\left(\tau\right),\\
	y_{n}\left(\tau\right) & =\sqrt{\frac{2}{\beta}}\sin\left(\frac{\pi n}{\beta}\tau\right).
	\end{split}
	\end{equation}
	Note that although $y\left(\beta\right)=y\left(0\right)=0$, its derivatives might jump between the two ends.
	Specifically for odd $n$ $y_n'\left(0\right)\ne y_n'\left(\beta\right)$.
	Performing the integrals gives
	\begin{equation}
	\begin{split}
	\bra{0}e^{-\beta H}\ket{0} & =\prod_{n=1}^{\infty}\int da_{n}\exp\left(-\frac{m}{2}\left(\left(\frac{\pi n}{\beta}\right)^{2}+\omega^{2}\right)a_{n}^{2}\right)\\
	& = \left(\prod_{n=1}^{\infty}\frac{m\pi n^{2}}{2\beta^{2}}\right)^{-\frac{1}{2}}\sqrt{\frac{\omega\beta}{\sinh\left(\omega\beta\right)}}.
	\end{split}
	\end{equation}
	Using the zeta-function regularization \eqref{eq:zeta_eq}
	$\prod_{n=1}^{\infty}\frac{m\pi n^{2}}{2\beta^{2}}=\frac{2\pi\beta}{\sqrt{\frac{m\pi}{2}}}$ and
	\begin{equation}
	\begin{split}
	\bra{0}e^{-\beta H}\ket{0}=\left(\frac{\pi}{2m}\right)^{\frac{1}{4}}\cdot\sqrt{\frac{m\omega}{2\pi\sinh\left(\omega\beta\right)}}.
	\end{split}
	\end{equation}
	We will ignore the $\left(\frac{\pi}{2m}\right)^{\frac{1}{4}}$ and take:
	\begin{equation}
	\begin{split}
	\bra{0}e^{-\beta H}\ket{0}=\sqrt{\frac{m\omega}{2\pi\sinh\left(\omega\beta\right)}}
	\end{split}
	\end{equation}
	Going back to \eqref{eq:K_E_y}, we find the on-shell action $S_{cl}=m\omega\left(\frac{\cosh\left(\omega\beta\right)-1}{\sinh\left(\omega\beta\right)}\right)x^{2}$, which gives the final propagator
	\begin{equation}\label{eq:KE_x}
	\bra{x}e^{-\beta H}\ket{x}=\sqrt{\frac{m\omega}{2\pi\sinh\left(\omega\beta\right)}}\exp\left(-m\omega\left(\frac{\cosh\left(\omega\beta\right)-1}{\sinh\left(\omega\beta\right)}\right)x^{2}\right)
	\end{equation}

	To find the partition-function \eqref{eq:HO_Z} we integrate over $x$, to get
	\begin{align}\label{eq:Z_QM}
	Z\left(\beta\right) & =\int dx \bra{x}e^{-\beta H}\ket{x}\nonumber \\
	& =\frac{1}{2\sinh\left(\frac{\omega\beta}{2}\right)}.
	\end{align}
	This result is the canonically normalized one, as we can see in the energy basis:
	\begin{equation}
	\begin{split}
	Z\left(\beta\right)=e^{-\frac{\omega\beta}{2}}\frac{1}{1-e^{-\omega\beta}}=\sum_{n=0}^{\infty}e^{-\omega\beta\left(n+\frac{1}{2}\right)}.
	\end{split}
	\end{equation}

	On the other hand, we can write the partition function \eqref{eq:HO_Z} formally as the path integral
	\begin{equation}
		Z(\beta)=\int Dx\left(\tau\right)\mid_{x\left(0\right)=x(\beta)} e^{-S\left[x\right]}
	\end{equation}
	Instead of the previous calculation, we would like to understand this expression as the path integral over the smooth manifold of circle $S^1_{\beta}$. In practice, we decompose $x(\tau)$ in the 
	Fourier basis (the fluctuations of the constant mode $x$ the $n=0$ mode):
	\begin{equation}
	\begin{split}
	x\left(\tau\right) & =\frac{1}{\sqrt{\beta}}b_{0}+\sum_{n=1}^{\infty}\left(b_{n}\sqrt{\frac{1}{2\beta}}\exp\left(i\frac{2\pi n}{\beta}\tau\right)+c.c.\right)
	\end{split}
	\end{equation}
	Note that unlike the previous decomposition \eqref{eq:y_basis}, this basis is smooth at $\tau=0,\beta$, and so naively might give different results. In this case the action is
	\begin{equation}
	\begin{split}
	S\left[x\right] & =\frac{m}{2}\omega^{2}b_{0}^{2}+\sum_{n=1}^{\infty}b_{n}^{*}b_{n}\frac{m}{2}\left(\left(\frac{2\pi n}{\beta}\right)^{2}+\omega^{2}\right).
	\end{split}
	\end{equation}
	Computing the Gaussian integral gives the partition function
	\begin{equation}
	\begin{split}
	Z\left(\beta\right) & =\left(\prod_{n=0}^{\infty}\int db_{n}\right)\exp\left(-S\right)\\
	 & =\sqrt{\frac{\pi}{\frac{m}{2}\omega^{2}}}\prod_{n=1}^{\infty}\frac{\pi}{\frac{m}{2}\left(\left(\frac{2\pi n}{\beta}\right)^{2}+\omega^{2}\right)}\\
	 & =\sqrt{\frac{2\pi}{m\omega^{2}}}\cdot\prod_{n=1}^{\infty}\frac{1}{\frac{2\pi m}{\beta^{2}}n^{2}}\cdot\frac{\frac{\omega\beta}{2}}{\sinh\left(\frac{\omega\beta}{2}\right)}.
	\end{split}
	\end{equation}
	The second term can be calculated using zeta function regularization
	\eqref{eq:zeta_eq} to give $\prod_{n=1}^{\infty}\frac{2\pi m}{\beta^{2}}n^{2}=\frac{2\pi}{\sqrt{\frac{2\pi m}{\beta^{2}}}}$, and
	\begin{align}
	Z\left(\beta\right) & =\sqrt{\frac{2\pi}{m\omega^{2}}\cdot}\frac{1}{2\pi}\sqrt{\frac{2\pi m}{\beta^{2}}\cdot}\frac{\frac{\omega\beta}{2}}{\sinh\left(\frac{\omega\beta}{2}\right)}\label{eq:Z_QM_smooth}\\
	 & =\frac{1}{2\sinh\left(\frac{\omega\beta}{2}\right)}.\nonumber 
	\end{align}
	As expected, this is the same result found using the discontinuous basis \eqref{eq:Z_QM}.

\subsubsection{Replica average}
	We turn to the more general case of $\overline{Z_1^N}$ (from \eqref{eq:av_renyi}). In one dimension and in the position space basis, this is the path integral on $N$ intervals with all their $2N$ ends identified to the same value $x$:
	\begin{equation}\label{eq:Z_N_QM_def}
	\overline{Z_1^N}=\int dx \bra{x}e^{-\beta H}\ket{x}^N.
	\end{equation}
	Using the expression for the propagation \eqref{eq:KE_x} we find
	\begin{equation}\label{eq:Z_N_QM}
	\begin{split}
	\overline{Z_1^N} & =\int dx\left(\frac{m\omega}{2\pi\sinh\left(\omega\beta\right)}\right)^{\frac{N}{2}}\exp\left(-N\cdot m\omega\left(\frac{\cosh\left(\omega\beta\right)-1}{\sinh\left(\omega\beta\right)}\right)x^{2}\right)\\
	 & = \frac{1}{\sqrt{N}}\left(\frac{m\omega}{2\pi\sinh\left(\omega\beta\right)}\right)^{\frac{N-1}{2}}\frac{1}{2\sinh\left(\frac{\omega\beta}{2}\right)}.
	\end{split}
	\end{equation}

\begin{figure}
		\centering
		\includegraphics[width=.7\linewidth]{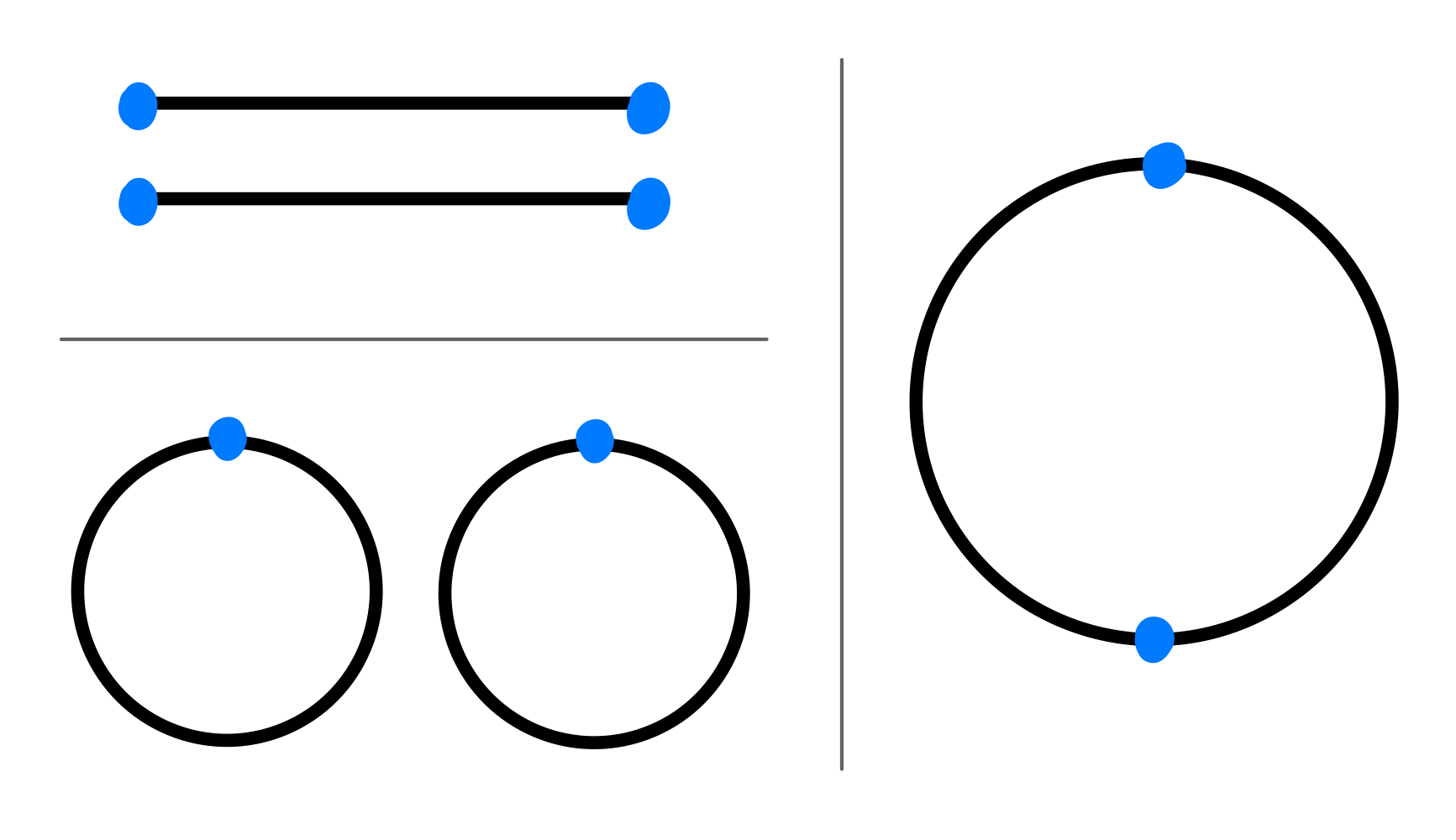}
		\caption{\label{fig:free_calc}
		The three equivalent path integrals. Top left: two intervals of size $\beta$ with all their ends identified (in blue). Bottom left: two circles of length $\beta$, with the point $\tau=0$ identified between them. Right: one circle of length $2\beta$, with the points $\tau=0,\beta$ further identified.
		}
	\end{figure}

	In the rest of this section, we focus on the $N=2$ case. Following the thermal calculation, the path integral \eqref{eq:Z_N_QM_def} can be equivalently calculated over two other topologies, which corresponds to two different smoothness assumptions at the crossing (see figure \eqref{fig:free_calc}). We show below that the path integrals on both topologies coincide with \eqref{eq:Z_N_QM}.

	\paragraph{Two circles}

	The first topology takes place by gluing each of the two intervals into two circles. We denote the particle on each of the circles by $x^{\left(1\right)}(\tau)$ and $x^{\left(2\right)}(\tau)$. On each circle we assume the particles are smooth, and share between them the value on $\tau=0$:
	\begin{equation}
	\begin{split}
	\overline{Z^2_1}=\int Dx^{\left(1\right)}\left(\tau\right)Dx^{\left(2\right)}\left(\tau\right) e^{-S[x]}
	\delta\left(x^{\left(1\right)}\left(0\right)-x^{\left(2\right)}\left(0\right)\right).
	\end{split}
	\end{equation}

	Decomposing the particles into real temporal modes
	\begin{equation}
	\begin{split}
	x^{\left(1\right)}\left(\tau\right) & =\frac{1}{\sqrt{\beta}}a_{0}^{\left(1\right)}+\sum_{n=1}^{\infty}\left(a_{n}^{\left(1\right)}\sqrt{\frac{2}{\beta}}\cos\left(\frac{2\pi n}{\beta}\tau\right)+b_{n}^{\left(1\right)}\sqrt{\frac{2}{\beta}}\sin\left(\frac{2\pi n}{\beta}\tau\right)\right)\\
	x^{\left(2\right)}\left(\tau\right) & =\frac{1}{\sqrt{\beta}}a_{0}^{\left(2\right)}+\sum_{n=1}^{\infty}\left(a_{n}^{\left(2\right)}\sqrt{\frac{2}{\beta}}\cos\left(\frac{2\pi n}{\beta}\tau\right)+b_{n}^{\left(2\right)}\sqrt{\frac{2}{\beta}}\sin\left(\frac{2\pi n}{\beta}\tau\right)\right)
	\end{split}
	\end{equation}
	we have
	\begin{equation}
	\begin{split}
	\overline{Z^2_1}=\prod_{i=1}^{2}\prod_{n=0}^{\infty}\int da_{n}^{\left(i\right)}db_{n}^{\left(i\right)}\exp\left(-S\right)\delta\left(x^{\left(1\right)}\left(0\right)-x^{\left(2\right)}\left(0\right)\right),
	\end{split}
	\end{equation}
	with the action
	\begin{equation}
	\begin{split}
	S & =\sum_{i=1}^{2}\frac{m}{2}\omega^{2}\left(a_{0}^{\left(i\right)}\right)^{2}+\sum_{n=1}^{\infty}\left(\left(a_{n}^{\left(i\right)}\right)^{2}+\left(b_{n}^{\left(i\right)}\right)^{2}\right)\frac{m}{2}\left(\left(\frac{2\pi n}{\beta}\right)^{2}+\omega^{2}\right).
	\end{split}
	\end{equation}
	In terms of the modes, the delta function constraint is
	\begin{equation}\label{eq:constraint}
		\frac{1}{\sqrt{\beta}}a_{0}^{\left(1\right)}+\sqrt{\frac{2}{\beta}}\sum_{n=1}^{\infty}a_{n}^{\left(1\right)}
		=\frac{1}{\sqrt{\beta}}a_{0}^{\left(2\right)}+\sqrt{\frac{2}{\beta}}\sum_{n=1}^{\infty}a_{n}^{\left(2\right)}/
	\end{equation}
	Note that
	the $b_{n}^{\left(i\right)}$s decouple and contribute (using \eqref{eq:Z_QM_smooth})
	\begin{align}\label{eq:Z_sin}
	Z_2^{sin} & =\prod_{n=1}^{\infty}\frac{\pi}{\frac{m}{2}\left(\left(\frac{2\pi n}{\beta}\right)^{2}+\omega^{2}\right)} =\sqrt{\frac{m}{2\pi}}\frac{\omega}{2\sinh\left(\frac{\omega\beta}{2}\right)}.
	\end{align}

	As for the $a_{n}^{\left(i\right)}$s, we can change basis to $a_{n}^{\pm} =\frac{1}{\sqrt{2}}\left(a_{n}^{\left(1\right)}\pm a_{n}^{\left(2\right)}\right)$. Integrating over $a_{0}^-$ using the delta-function \eqref{eq:constraint} gives an action
	\begin{align}\label{eq:action_change_basis}
	S & =\sum_{n=0}^{\infty}\left(a_{n}^{+}\right)^{2}\frac{m}{2}\left(\left(\frac{2\pi n}{\beta}\right)^{2}+\omega^{2}\right)\nonumber \\
	 & +\sum_{n=1}^{\infty}\left(a_{n}^{-}\right)^{2}\frac{m}{2}\left(\left(\frac{2\pi n}{\beta}\right)^{2}+\omega^{2}\right)+m\omega^{2}\left(\sum_{n=1}^{\infty}a_{n}^{-}\right)^{2}.
	\end{align}
	The $a_{n}^{+}$s decouple as well, and contribute
	\begin{equation}
	Z_{2}^{+}=\prod_{n=1}^{\infty}\sqrt{\frac{\pi}{\frac{m}{2}\left(\left(\frac{2\pi n}{\beta}\right)^{2}+\omega^{2}\right)}}=\left(\frac{1}{\omega}
		\sqrt{\frac{2\pi}{m}}\frac{1}{2\sinh\left(\frac{\omega\beta}{2}\right)}\right)^{\frac{1}{2}}.\label{eq:Z_plus}
	\end{equation}
	In order to carry the Gaussian integral over the $a_n^-$s, we need the determinant of the quadratic form \eqref{eq:action_change_basis}. Consider first the matrix $M\left(\mu\right)_{n,m}=\lambda_{n}\delta_{n,m}+\mu$. Its determinant satisfy
	\begin{equation}\label{eq:matrix_identity}
	\begin{split}
	\frac{\det M\left(\mu\right)}{\det M\left(0\right)}=1+\mu\sum_{n=0}^{\infty}\lambda_{n}^{-1}.
	\end{split}
	\end{equation}
	Comparing to the remaining $a_{n}^{-}$ action \eqref{eq:action_change_basis} we find
	$\lambda_{n}=\frac{m}{2}\left(\left(\frac{2\pi n}{\beta}\right)^{2}+\omega^{2}\right)$ 
	and $\mu= m\omega^{2}$. The resulted integral over $a_n^-$ is
	\begin{equation}
	\begin{split}
	Z^{-}_2 & =\sqrt{\left(\prod_{n=1}^{\infty}\frac{\pi}{\frac{m}{2}\left(\left(\frac{2\pi n}{\beta}\right)^{2}+\omega^{2}\right)}\right)\cdot\left(1+m\omega^{2}\left(\sum_{n=1}^{\infty}\frac{1}{\frac{m}{2}\left(\left(\frac{2\pi n}{\beta}\right)^{2}+\omega^{2}\right)}\right)\right)^{-1}}\\
	 & =\sqrt{\sqrt{\frac{m}{2\pi}}\frac{1}{\beta\cosh\left(\frac{\omega\beta}{2}\right)}}.
	\end{split}
	\end{equation}
	In the second line we used the identity $\sum_{n=1}^\infty \frac{1}{\left(\frac{\pi n}{a}\right)^2+1} = \frac{1}{2}\left(a\coth(a)-1\right)$. Together with \eqref{eq:Z_sin} and \eqref{eq:Z_plus} we find (we also need to multiply
	by $\sqrt{\beta/2}$ from the delta function for $a_{0}^{\left(2\right)}$
	 \eqref{eq:constraint}):
	\begin{equation}
	\begin{split}
	\overline{Z^2_1} & = \sqrt{\frac{\beta}{2}} \cdot Z_2^{sin} \cdot Z_2^+ \cdot Z_2^{-}\\
	 & =\frac{1}{\sqrt{2}}\frac{1}{2\sinh\left(\frac{\omega\beta}{2}\right)}\cdot\sqrt{\frac{1}{2\pi}\frac{m\omega}{\sinh\left(\omega\beta\right)}}
	\end{split}
	\end{equation}
	This is exactly \eqref{eq:Z_N_QM} for $N=2$.

	\paragraph{One big circle}
	The second topology we can consider for $\overline{Z^2_1}$ is built by connecting the two intervals together into one circle of size $2\beta$, $x(0)=x(2\beta)$, with a delta-function identification $x(0)=x(\beta)$:
	\begin{equation}
	\begin{split}
	\overline{Z^2_1}=\int Dx\left(\tau\right)e^{-S[x]}\delta\left(x\left(0\right)-x\left(\beta\right)\right).
	\end{split}
	\end{equation}
	Decomposing $x(\tau)$ in the modes of the $2\beta$ circle
	\begin{equation}
	\begin{split}
	x\left(\tau\right) & =\frac{1}{\sqrt{2\beta}}a_{0}
	+\sum_{n=1}^{\infty}\left(a_{n}\sqrt{\frac{1}{\beta}}\cos\left(\frac{\pi n}{\beta}\tau\right)+b_{n}\sqrt{\frac{1}{\beta}}\sin\left(\frac{\pi n}{\beta}\tau\right)\right)
	\end{split}
	\end{equation}
	gives the integral
	\begin{equation}
	\begin{split}
	\overline{Z^2_1}=\prod_{n=0}^{\infty}\int da_{n}db_{n}\exp\left(-S\right)\delta\left(x\left(0\right)-x\left(\beta\right)\right),
	\end{split}
	\end{equation}
	with the action
	\begin{equation}
	\begin{split}
	S & =\frac{m}{2}\omega^{2}a_{0}^{2}
	+\sum_{n=1}^{\infty}\left(a_{n}^{2}+b_{n}^{2}\right)\frac{m}{2}\left(\left(\frac{\pi n}{\beta}\right)^{2}+\omega^{2}\right).
	\end{split}
	\end{equation}
	The delta function gives the constraint
	\begin{equation}
	2\sqrt{\frac{1}{\beta}}\sum_{n=0}^{\infty}a_{2n+1}=0.\label{eq:constraint_2}
	\end{equation}
	Both the $b_n$s and the even $a_{2n}$ modes decouple from the constraint and contribute
	\begin{equation}\label{eq:Zb_and_Zeven}
	\begin{split}
		Z_2^b &= \prod_{n=0}^{\infty}\sqrt{\frac{\pi}{\frac{m}{2}\left(\left(\frac{\pi n}{\beta}\right)^{2}+\omega^{2}\right)}}=\left(\frac{1}{\omega}
		\sqrt{\frac{2\pi}{m}}\frac{1}{2\sinh\left(\omega\beta\right)}\right)^{\frac{1}{2}},\\
		Z_2^{even} & = \prod_{n=1}^{\infty}\sqrt{\frac{\pi}{\frac{m}{2}\left(\left(\frac{2\pi n}{\beta}\right)^{2}+\omega^{2}\right)}}=\left(\sqrt{\frac{m}{2\pi}}\frac{\omega}{2\sinh\left(\frac{\omega\beta}{2}\right)}\right)^{\frac{1}{2}}.
	\end{split}
	\end{equation}
	To deal with the odd $a_n$s we denote $c_n = a_{2n+1}$. Integrating over $c_{0}$ using the delta-function \eqref{eq:constraint_2} gives $c_0 = -\sum_{n=1}^\infty c_n$. We get following quadratic action for the remaining $c_n$ $n=1,2,...$:
	\begin{equation}
	\begin{split}
	S & = \sum_{n=1}^{\infty} c_{n}^{2}\frac{m}{2}\left(\left(\frac{\pi (2n+1)}{\beta}\right)^{2}+\omega^{2}\right) + \sum_{n,m=1}^{\infty} c_{n}c_m \frac{m}{2}\left(\left(\frac{\pi}{\beta}\right)^{2}+\omega^{2}\right).
	\end{split}
	\end{equation}
	The quadratic form here is exactly of the type \eqref{eq:matrix_identity}, this time with $\lambda_n = \frac{m}{2}\left(\left(\frac{\pi (2n+1)}{\beta}\right)^{2}+\omega^{2}\right)$ and $\mu = \frac{m}{2}\left(\left(\frac{\pi}{\beta}\right)^{2}+\omega^{2}\right)$. Therefore 
	\begin{equation}\label{eq:Z_odd}
	\begin{split}
		Z_2^{odd} & = \sqrt{\prod_{n=1}^{\infty}\frac{\pi}{\frac{m}{2}\left(\left(\frac{\pi (2n+1)}{\beta}\right)^{2}+\omega^{2}\right)} 
		\cdot \left( 1+ \sum_{n=1}^\infty \frac{\left(\frac{\pi}{\beta}\right)^{2}+\omega^{2}}{\left(\frac{\pi (2n+1)}{\beta}\right)^{2}+\omega^{2}} \right)^{-1}} \\
		& = \sqrt{\frac{m\omega}{\pi\beta}\frac{1}{\sinh\left(\frac{\omega\beta}{2}\right)}}
	\end{split}
	\end{equation}
	Together \eqref{eq:Zb_and_Zeven}, \eqref{eq:Z_odd} and the factor of $\frac{\sqrt{\beta}}{2}$ from \eqref{eq:constraint_2} gives
	\begin{equation}
	\begin{split}
		Z_2 &= \frac{\sqrt{\beta}}{2}\cdot Z_2^b \cdot Z_2^{even} \cdot Z_2^{odd} \\
		& = \frac{1}{\sqrt{2}}\frac{1}{2\sinh\left(\frac{\omega\beta}{2}\right)}\cdot\sqrt{\frac{1}{2\pi}\frac{m\omega}{\sinh\left(\omega\beta\right)}}
	\end{split}
	\end{equation}
	Again, this is exactly equal to the canonical result \eqref{eq:Z_N_QM}.

\subsection{Free theory on a circle}
In this section, we consider the two-dimensional field theory of a single free scalar $\phi(x)$ of mass $m$, on a circle of size $L$, $x\sim x+L$. The Euclidean action is
\begin{equation}
	S\left[\phi\right]=\int_{0}^{L}dx\int_{0}^{\beta}d\tau\frac{1}{2}\left(\partial_{i}\phi\partial^{i}\phi+m^{2}\phi^{2}\right).
\end{equation}
We start by studying the conventional thermal partition function. We show explicitly that it can be calculated either as the path integral on the identified cylinder, or as the path integral on the torus. We then turn to the field theory calculation of $\overline{Z_1^N}$ and discuss its divergences.
\subsubsection{Thermal partition-function}
	Using the path integral formalism, the thermal partition function can be written as a path integral over the cylinder $S^1_L\times [0,\beta]$, with the two ends identified
	\begin{equation}\label{eq:Z_free_def}
	\begin{split}
	Z\left(\beta\right) & =\text{Tr}\left(e^{-\beta H}\right)\\
	 & =\int D\tilde{\phi}\left(x\right)\bra{\tilde{\phi}}e^{-\beta H}\ket{\tilde{\phi}}\\
	 & =\int D\tilde{\phi}\left(x\right)\int D\phi\left(x,\tau\right)\mid_{\phi\left(x,0\right)=\tilde{\phi}}^{\phi\left(x,\beta\right)=\tilde{\phi}}e^{-S}
	\end{split}
	\end{equation}
	As a first step, we decompose $\phi\left(x,\tau\right)$ into spatial modes
	\begin{equation}
	\begin{split}
	\phi\left(x,\tau\right) & =\sum_{n=0}^{\infty}\phi_{n}\left(\tau\right)\sqrt{\frac{1}{2L}}e^{i\frac{2\pi n}{L}x}+c.c.\\
	\tilde{\phi}\left(x\right) & =\sum_{n=0}^{\infty}\tilde{\phi}_{n}\sqrt{\frac{1}{2L}}e^{i\frac{2\pi n}{L}x}+c.c.
	\end{split}
	\end{equation}
	We can write the path integral as
	\begin{equation}
	\begin{split}
	Z\left(\beta\right) =&\int d\tilde{\phi}_{0}\int D\phi_{0}\mid_{\phi_{0}\left(0\right)=\tilde{\phi}_{0}}^{\phi_{0}\left(\beta\right)=\tilde{\phi}_{0}}\\
	&\prod_{n=1}^{\infty}\int d\tilde{\phi}_{n}d\tilde{\phi}_{n}^{*}\int D\phi_{n}D\phi_{n}^{*}\mid_{\phi_{n}\left(0\right)=\tilde{\phi}_{n}}^{\phi_{n}\left(\beta\right)=\tilde{\phi}_{n}}\exp\left(-S\left[\phi\right]\right),
	\end{split}
	\end{equation}
	with the action
	\begin{equation}
	\begin{split}
	S\left[\phi\right]=\sum_{n=0}^{\infty}\int_{0}^{\beta}d\tau\frac{1}{2}\left(\dot{\phi}_{n}^{*}\dot{\phi}_{n}+\left(\left(\frac{2\pi n}{L}\right)^{2}+m^{2}\right)\phi_{n}^{*}\phi_{n}\right).
	\end{split}
	\end{equation}
	Each mode $\phi_n(\tau)$ is a complex harmonic oscillator with $\omega_n^2=\left(\frac{2\pi n}{L}\right)^{2}+m^{2}$ (besides $n=0$ which is real with $\omega_0=m$). The path integral on each mode is over an identified interval. In the previous section, we found that this path integral is also equal to the one over the circle. Composing back the spatial modes, we learn \eqref{eq:Z_free_def} is equal to the free-theory path integral over the torus $S^1_L\times S^1_\beta$! 
	We can find its value using \eqref{eq:Z_QM}
	\begin{equation}\label{eq:1d_Z_final}
	\begin{split}
	Z\left(\beta\right) & =\frac{1}{2\sinh\left(\frac{\beta m}{2}\right)}\prod_{n=1}^{\infty}\frac{1}{\left(2\sinh\left(\frac{\beta}{2}\sqrt{\left(\frac{2\pi n}{L}\right)^{2}+m^{2}}\right)\right)^{2}}\\
	& = \frac{e^{-\beta\tilde{E}_{0}}}{1-e^{-\beta m}}\left(\prod_{n=1}^{\infty}\frac{1}{1-\exp\left(-\beta\sqrt{\left(\frac{2\pi n}{L}\right)^{2}+m^{2}}\right)}\right)^2,
	\end{split}
	\end{equation}
	With $\tilde{E}_{0}$ being the regularization of the zero-point energy $E_{0}=\frac{m}{2} + \sum_{n=1}^{\infty}\sqrt{\left(\frac{2\pi n}{L}\right)^{2}+m^{2}}$.
	For $m=0$ (and ignoring the zero mode), we can use the standard $\sum_{n=1}^{\infty}n=-\frac{1}{12}$ to get the known result
	\begin{equation}
	\begin{split}
	Z\left(\beta\right) =\left(e^{\frac{\pi}{12}\frac{\beta}{L}}\prod_{n=1}^{\infty}\frac{1}{1-\exp\left(-\frac{2\pi\beta}{L}n\right)}\right)^{2} =\frac{1}{\eta^{2}\left(i\frac{\beta}{L}\right)}.
	\end{split}
	\end{equation}
	For the sake of the next section, we can also find $Z(\beta)$ differently. Already in \eqref{eq:Z_free_def} we can first evaluate $\bra{\tilde{\phi}}e^{-\beta H}\ket{\tilde{\phi}}$.
	By decomposing $\tilde \phi$ into modes and regularizing the zero-point energy in the same way we get, using \eqref{eq:KE_x},
	\begin{equation}\label{eq:phi_prop_ren}
	\begin{split}
		\bra{\tilde{\phi}}e^{-\beta H}\ket{\tilde{\phi}} &= e^{-\beta \tilde E_0}
		\sqrt{\frac{m}{\pi(1-e^{-2 \beta m})}}\exp\left(-2m\frac{\sinh^2\left(\frac{\beta m}{2}\right)}{\sinh\left(\beta m\right)}\tilde\phi_0^{2}\right)\\
		&\times \prod_{n=1}^\infty 
		\frac{\omega_n}{\pi(1-e^{-2\beta\omega_n})}
		\exp\left(-4\omega_n\frac{\sinh^2\left(\frac{\beta\omega_n}{2}\right)}{\sinh\left(\beta\omega_n\right)}\tilde\phi_n^* \tilde\phi_n \right).
	\end{split}
	\end{equation}
	This expression was renormalized only using cylinder local counter-terms. In particular, integrating over $D\tilde\phi$ will give back the final thermal partition-function \eqref{eq:1d_Z_final}. It agrees with our expectation that the latter can be regularized using only local counter-terms on the torus.

\subsubsection{Replica average}
	We generalized the discussion to the free-theory calculation of $\overline{Z_1^N}$ in the field basis (\eqref{eq:av_renyi})
	\begin{equation}
	\begin{split}
	\overline{Z_1^N} & =\int D\tilde{\phi}\left(x\right)\left(\int D\phi\left(x,\tau\right)\mid_{\phi\left(x,0\right)=\tilde{\phi}}^{\phi\left(x,\beta\right)=\tilde{\phi}}\exp\left(-S\right)\right)^{N}\\
	&=\int D\tilde{\phi}\left(x\right)\left(\bra{\tilde{\phi}}e^{-\beta H}\ket{\tilde{\phi}}\right)^{N}.
	\end{split}
	\end{equation}
	Using \eqref{eq:phi_prop_ren} and integrating the $\tilde \phi$ modes gives
	\begin{equation}
	\begin{split}
		\overline{Z_1^N} & =\int D\tilde{\phi}\left(x\right)\left(\bra{\tilde{\phi}}e^{-\beta H}\ket{\tilde{\phi}}\right)^{N}\\
		& =\frac{e^{-N\beta\tilde E_0}}{1-e^{- \beta m }}
		\frac{1}{\sqrt{N}}\left(\frac{m}{\pi(1-e^{-2\beta m})}\right)^{\frac{N-1}{2}}
		\\
		&\cdot\prod_{n=1}^{\infty}\frac{1}{N}\left(\frac{\omega_n}{\pi\left(1-e^{-2\beta\omega_n}\right)}\right)^{N-1}\cdot\frac{1}{\left(1-e^{-\beta \omega_n }\right)^{2}}.
	\end{split}
	\end{equation}
	Remember that \eqref{eq:phi_prop_ren} was already regulated using local counter-terms on the cylinder, which gave finite thermal partition function on \eqref{eq:1d_Z_final}. Still, for every $N > 1$ we have a divergence due to $\prod_{n=1}^\infty \frac{\omega_n^{N-1}}{N \pi^{N-1}}$. 

	More generally, for free theories on $S^{d-1}_{R}\times S^1$ the product would be on the sphere's Laplacian eigenvalues $\prod_{n=0}^\infty \sqrt{\frac{\lambda_n^{N-1}}{N \pi^{N-1}}}$. Using a momentum cutoff $\Lambda$ and a dimensionless path integral measure  \cite{Anninos:2012ft} gives a free-energy contribution of $(N-1)\frac{1}{2}\sum_{n=0}^{\Lambda R} d_n \log(\lambda_n/\Lambda)$ which behave at large $\Lambda$ as $\sim (N-1)(R\Lambda)^{d-1}$. Thus this divergence can be canceled by a local (note the $\beta$ independence) counter-term on the $N-1$ spheres we identified. We expect that also in weak coupling, two kinds of counter terms are necessary: the local counter-terms on the $N$ cylinders used also for the thermal $N=1$ calculation, and counter-terms localized to the $N-1$ identified circles to account for the divergences of the identification.
	
	We go back to the two-dimensional case and renormalized the divergence using zeta-function-regularization \eqref{eq:zeta_eq}, $\prod_{n=1}^{\infty}\frac{1}{N(L/2)^{N-1}}n^{N-1}=\sqrt{N\left(\pi L\right)^{N-1}}$ we have
	\begin{equation}
	\begin{split}
	\overline{Z_1^N} & =e^{-N\beta\tilde E_0} \ \left(\frac{mL}{1-e^{-2\beta m}}\right)^{N-1} \frac{1}{1-e^{-\beta m}} \cdot \prod_{n=1}^\infty \left(\frac{\sqrt{1+\left(\frac{m L}{2\pi n}\right)^2}}{1-e^{-2\beta\omega_n}}\right)^{N-1} \frac{1}{\left(1-e^{-\beta \omega_n}\right)^2}\\
	\end{split}
	\end{equation}
	At $m=0$ (and removing the zero mode) we can simplify the expression using $\prod_{n=1}^{\infty}\frac{1}{1-e^{-an}}=\frac{1}{e^{\frac{a}{24}}\eta\left(\frac{a}{2\pi}i\right)}$
	\begin{equation}
	\overline{Z_1^N}=\sqrt{N}\left(\frac{2\pi\sinh\left(\frac{mL}{2}\right)}{m \ \eta^{2}\left(i\frac{2\beta}{L}\right)}\right)^{\frac{N-1}{2}}\frac{1}{\eta^{2}\left(i\frac{\beta}{L}\right)}.
	\label{eq:Z_N_circle}
	\end{equation}

	As a comparison note that in the energy basis
	\begin{equation}
		\overline{Z^N_1} = \sum_n \bra{n} \exp(-\beta H) \ket{n}^N = Z(\beta N) = \eta^{-2}\left(i\frac{\beta N}{L}\right)
	\end{equation}
	Importantly $\bra{n} \exp(-\beta H) \ket{n}$ has only ground state divergence $e^{-\tilde E_0 \beta}$ and did not seem to require the localized counter-terms we introduced above. We learn that the divergences of the identification are basis-dependent.

\bibliographystyle{JHEP}
\bibliography{ref}
\end{document}